\documentclass[12pt]{article}
\usepackage{amsmath}         % \
\usepackage{amssymb}         %  | AMS Packages for math
\usepackage{amsfonts}        % /
\usepackage{graphicx}        % Graphics
\usepackage{booktabs}

\usepackage{lipsum}        % block of text (formatting test)
\usepackage{framed}        % boxed remarks
\usepackage{subcaption}    % subfigures; subfig depreciated
\usepackage{multirow}      % multiple row elements in a table
\usepackage{cite}          % grouped cites (incompat. w/ collref)
\usepackage{arydshln} 	    % dashed lines in arrays and tables
\usepackage{bbm}           % \mathbbm{1} incompatible with XeLaTeX 
\usepackage{tocloft}       % Table of Contents	
\usepackage[normalem]{ulem} % for \sout
\usepackage{fancyhdr}		% to put preprint number
\usepackage[utf8]{inputenc}
\usepackage{scalefnt}		% for definition of \smaller
\usepackage{xspace}			% for spacing in macros
\usepackage{pifont}			% for 'x' dingbat 

\usepackage{multicol}
\usepackage{etoolbox}
\usepackage{relsize}
\patchcmd{\thebibliography}
  {\list}
  {\begin{multicols}{2}\smaller\list}
  {}
  {}
\appto{\endthebibliography}{\end{multicols}}

\usepackage{tikzfeynman}   % Flip's Feynman Diagrams

\usepackage[margin=2cm]{geometry}   % reasonable margins
\graphicspath{{figures/}}	        % set directory for figures
\numberwithin{equation}{section}    % set equation numbering
\renewcommand{\vec}[1]{\mathbf{#1}} % vectors are boldface

\newcommand{\acro}[1]{\textsc{\MakeLowercase{#1}}\xspace}    
\newcommand{\email}[1]{\href{mailto:#1}{#1}}

\newenvironment{institutions}[1][2em]{\begin{list}{}{\setlength\leftmargin{#1}\setlength\rightmargin{#1}}\item[]}{\end{list}}

\fancypagestyle{firststyle}
{
	\rhead{\footnotesize \texttt{UCI-TR-2016-21}}
   \fancyfoot{}
}

\usepackage{etoc}
\renewcommand{\etocaftertitlehook}{\pagestyle{plain}}
\renewcommand{\etocaftertochook}{\thispagestyle{firststyle}}

\usepackage[
	colorlinks=true,
	citecolor=black,
	linkcolor=black,
	urlcolor=blue,
	hypertexnames=false]{hyperref}

\begin{document}

\thispagestyle{firststyle}

\begin{center}

    {\huge \bf  Lepton-Flavor Violating Mediators}

    \vskip .7cm

	{ \textbf{Iftah Galon}$^a$, 
	  \textbf{Anna Kwa}$^a$, 
	  and \textbf{Philip Tanedo}$^{b}$
	  } 
    \\ \vspace{-.2em}
    { \tt
    \footnotesize
    \email{iftachg@uci.edu}, 
    \email{akwa@uci.edu},
    \email{flip.tanedo@ucr.edu} 
    }
	
    \vspace{-.2cm}
%	\vspace{-.3cm}

    \begin{institutions}[2.25cm]
    \footnotesize
    $^{a}$ {\it 
        Department of Physics \& Astronomy, 
	    University of California, 
	    Irvine, {CA} 92697
        }
	\\ 
	\vspace*{0.05cm}
	$^{b}$
	{\it 
	    Department of Physics \& Astronomy, University of  California, Riverside, {CA} 92521
	    }   
    \end{institutions}

\end{center}

%%%%%%%%%%%%%%%%%%%%%
%%%  ABSTRACT    %%%%
%%%%%%%%%%%%%%%%%%%%%

\begin{abstract}
\noindent 
We present a framework where dark matter interacts with the Standard Model through a light, spin-0 mediator that couples chirally to pairs of different-flavor leptons. 
This flavor violating final state weakens bounds on new physics coupled to leptons from terrestrial experiments and cosmic-ray measurements. 
As an example, we apply this framework to construct a model for the Fermi-LAT excess of GeV $\gamma$-rays from the galactic center.
We comment on the viability of this portal for self-interacting dark matter explanations of small scale structure anomalies and embeddings in flavor models. 
Models of this type are shown to be compatible with the muon anomalous magnetic moment anomaly. We review current experimental constraints and identify possible future theoretical and experimental directions.
\end{abstract}

\small
\setcounter{tocdepth}{2}
\tableofcontents
\normalsize
\clearpage

\section{Introduction}

Scenarios where dark matter is a thermal relic that interacts directly with the Standard Model are typically constrained by a range of complementary experimental searches~\cite{Arrenberg:2013rzp}. 
On the other hand, if dark matter is secluded from the Standard Model and only interacts through a light mediator, then one may obtain the observed relic density from thermal freeze out while parametrically suppressing signatures in direct detection and collider experiments~\cite{Pospelov:2007mp}. 
Direct annihilation into on-shell mediators sets the dark matter--mediator couplings, while the mediator--Standard Model couplings can be small enough to explain the null results from direct detection and collider experiments.
Dark matter continues to annihilate in the present day and the  Standard Model byproducts of the ensuing mediator decays may be observable.

One possible signature consistent with this framework is the recent excess of $\gamma$-rays from the Galactic Center observed by independent analyses of the Fermi Large Area Telescope (\acro{LAT}) data~\cite{
Goodenough:2009gk, 
Hooper:2010mq, 
Abazajian:2010zy, 
Boyarsky:2010dr, 
Hooper:2011ti, 
Abazajian:2012pn, 
Gordon:2013vta, 
Hooper:2013rwa, 
Huang:2013pda, 
Okada:2013bna, 
Macias:2013vya, 
Abazajian:2014fta, 
Daylan:2014rsa, 
Zhou:2014lva, 
Calore:2014xka, 
Calore:2014nla, 
Calore:2015nua} and directly by the Fermi-\acro{LAT} collaboration~\cite{TheFermi-LAT:2015kwa}. Alternative explanations include unresolved pulsars~\cite{Abazajian:2010zy, Abazajian:2012pn, Hooper:2013nhl, Abazajian:2014fta, Yuan:2014rca, Petrovic:2014xra, Bartels:2015aea, Lee:2015fea, McDermott:2015ydv, Hooper:2015jlu} or cosmic ray outbursts~\cite{Petrovic:2014uda, Carlson:2014cwa, Carlson:2015daa, Cholis:2015dea}.
The excess can be fit to effective theories that describe the annihilation of dark matter into pairs of Standard Model particles~\cite{Boehm:2014hva,Alves:2014yha,Ipek:2014gua,Izaguirre:2014vva,Berlin:2014tja,Agrawal:2014oha}.
Intriguingly, the total flux of excess $\gamma$-rays is consistent with the $s$-wave dark matter annihilation cross-section required to produce the observed relic density after thermal freeze out. Early fits to the energy spectrum preferred $40~\text{GeV}$ dark matter annihilating to $b$-quarks or $10~\text{GeV}$ dark matter annihilating into $\tau$-leptons; however, later studies found that masses up to $\mathcal O(100~\text{GeV})$ and a range of final states are allowed after accounting for the systematic uncertainties in the astrophysical background models~\cite{Zhou:2014lva, Calore:2014xka, Calore:2014nla,Agrawal:2014oha, Calore:2015nua, TheFermi-LAT:2015kwa}.
When dark matter annihilates into on-shell mediators in the secluded dark matter framework, the target space is shifted towards heavier dark matter and larger annihilation cross-sections~\cite{Boehm:2014bia, Abdullah:2014lla, Martin:2014sxa, Rajaraman:2015xka, Elor:2015tva}.

Most proposals to explain the excess from dark matter annihilations focus on $\gamma$-ray emission from bremsstrahlung and $\pi^0$ decays of annihilation products. 
These processes produce prompt $\gamma$-rays at the site of annihilation with intensities directly proportional to the square of the dark matter density. 
This predicts a similar signal in dwarf spheroidal galaxies which are rich in dark matter and have simpler astrophysical backgrounds than the Galactic Center~\cite{Abazajian:2015raa}. The non-observation of a clear excess in dwarf spheroidal galaxies~\cite{Drlica-Wagner:2015xua,Ackermann:2015zua} is typically a tension in dark matter interpretations of the $\gamma$-ray excess, indications of a possible excess in Reticulum II~\cite{Geringer-Sameth:2015lua,Hooper:2015ula} notwithstanding.

Lacroix, Boehm, and Silk pointed out that another mechanism by which the Galactic Center excess might be generated is through the inverse Compton scattering (ICS) of final-state electrons and positrons with infrared starlight~\cite{Lacroix:2014eea}. The energetic $e^+e^-$ pairs up-scatter the low-energy photons into the GeV range.
Recently, Calore et al.\ and Kaplinghat et al.\ proposed the possibility that these electron pairs may result from the decay of on-shell mediators~\cite{Calore:2014nla, Kaplinghat:2015gha}.
Up-scattering of starlight effectively does not occur in dwarf galaxies because of their much weaker interstellar radiation field. This removes the tension between the Fermi Galactic Center and dwarf $\gamma$-ray observations.
In its simplest form, however, this scenario is in tension with a different astrophysical observation.
Direct dark matter annihilation into $e^+e^-$ pairs produces a line in the local $e^+e^-$ spectrum that is observable by the Alpha Magnetic Spectrometer (AMS-02) telescope. The absence of such a line requires a mechanism to soften the primary $e^+e^-$ spectrum. Kaplinghat, Linden, and Yu realize this in the secluded dark matter scenario in which the annihilations into light mediators broadens the spectrum of daughter electrons and positrons.
When there is a hierarchy in the dark matter $\chi$ and mediator $\varphi$ masses, $m_\chi \gg m_\varphi \gg m_e$, the production spectrum of electrons is smeared from a line at $m_\chi$ to a box from 0 to $m_\chi$. In this way, one may attempt to hide the electron--positron spectrum by smearing it out within the AMS-02 error bars. The benchmark model in~\cite{Kaplinghat:2015gha} invokes 50~GeV dark matter annihilating into 100~MeV spin-1 mediators. As an additional feature, these masses automatically furnish the ingredients for a self-interacting dark matter solution to outstanding small-scale structure anomalies as reviewed in~\cite{Tulin:2013teo, Kaplinghat:2015aga}. The resulting dark matter annihilation cross-section is smaller than that of a thermal relic in the visible sector and thus requires the dark sector temperature to differ from the Standard Model at freeze out.

In this manuscript we introduce an alternative class of models that produce a $\gamma$-ray signal in the galactic center while avoiding bounds from the non-observation of such a signal in dwarf galaxies. Like Calore et al.\ and Kaplinghat et al., dark matter annihilates into on-shell mediators and the $\gamma$-ray signal is produced from the up-scattering of starlight.
In our case, however, the mediator is a spin-0 particle which decays into different-flavor lepton pairs, $\tau\mu$, $\tau e$ or $\mu e$. The final $e^\pm$ spectrum is softer which allows it to better fit within the error bars of the AMS-02 observations.
In the models with $\tau$ couplings, the hadronic $\tau$ decays yield prompt photons which contribute to the $\gamma$-ray excess in the absence of an interstellar radiation field. This, in turn, re-introduces tension with the dwarfs.
As observed in \cite{Abdullah:2014lla, Rajaraman:2015xka}, the leading $s$-wave contribution to dark matter annihilation into pseudoscalars is the $2\to 3$ process that further softens the spectrum of ensuing the Standard Model decay products. We show that the resulting $\gamma$-ray spectra are consistent with the Fermi-{LAT} excess for an annihilation cross-section that is compatible with the correct dark matter relic density even when the dark sector and Standard Model have the same temperature at freeze out. In the case where the mediator also contains a parity-even spin-$0$ component, this framework may still furnish a self-interacting dark matter solution to small-scale structure anomalies.

New sources of lepton flavor violation are strongly constrained by a plethora of flavor observables. 
Though we include a flavor-violating coupling, 
our scenario avoids many of these constraints by
preserving a residual $L_i - L_j$ global Abelian symmetry, under which the mediator has charge 2.
This restricts mediator--SM interactions to:
(a) a chiral coupling to a single oppositely-charged, different-flavored lepton pair, and (b) scalar potential terms proportional to powers of $\varphi^*\varphi$.
The symmetry prohibits most charged lepton flavor-violating processes while the chiral structure suppresses contributions to flavor-diagonal observables such as anomalous dipole moments. The leading constraints on the mediator's leptonic coupling come from the forward--backward asymmetries in electron collisions and from the anomalous magnetic moment of the muon. We identify possible future experimental directions in dark photon and collider searches.

The proposed coupling structure can naturally arise at the electroweak scale if a theory of flavor is responsible for the structure of both the Standard Model and mediator leptonic couplings. One example is to use the Froggatt--Nielsen mechanism~\cite{Froggatt:1978nt} and break its flavor symmetry at $\sim \rm{TeV}$. While this coupling structure is not stable under renormalization group evolution, loop corrections cannot 
substantially reintroduce additional flavor combinations of mediator leptonic couplings.
This phenomenon is due to the $L_i-L_j$ symmetry breaking spurions being proportional to the neutrino masses and the small electroweak-scale $\varphi$ couplings.

Recently, light, lepton flavor violating bosons have been a topic of interest~\cite{Heeck:2016xkh, Altmannshofer:2016brv}. 
Part of the interest has been driven by a possible breakdown of lepton flavor universality in $B$-meson decays~\cite{Aaij:2014ora,Lees:2012xj,Huschle:2015rga,Abdesselam:2016cgx,Aaij:2015yra} and early suggestions of the observation of the lepton-flavor violating Higgs decay, $h\to\mu\tau$~\cite{Khachatryan:2015kon, Aad:2015gha}. 
We point out that independent of the connection to possible astrophysical signals, the features of the lepton flavor violating interactions here are an interesting example for light, weakly coupled new physics which may have non-trivial signatures while evading existing bounds.
In addition, while our models are benchmarked against the Fermi-{LAT} $\gamma$-ray excess, they constitute a new class of simplified dark matter models where the mediator interacts with the Standard Model through chiral, lepton-flavor violating couplings. This fits into a larger body of literature studying dark sectors with non-trivial properties under the Standard Model flavor symmetries~\cite{
Kile:2011mn, 
Batell:2011tc, 
Kamenik:2011nb, 
Agrawal:2011ze, 
Lopez-Honorez:2013wla, 
Batell:2013zwa, 
Kile:2013ola, 
Kumar:2013hfa, 
Agrawal:2014una, 
Agrawal:2014aoa, 
D'Hondt:2015jbs, 
Agrawal:2015kje, 
Chen:2015jkt, 
Bhattacharya:2015xha, 
Bishara:2015mha, 
Agrawal:2015tfa, 
Haisch:2015ioa, 
Calibbi:2015sfa, 
Kilic:2015vka}.

\section{Models of a Lepton-Flavor Violating Mediator}

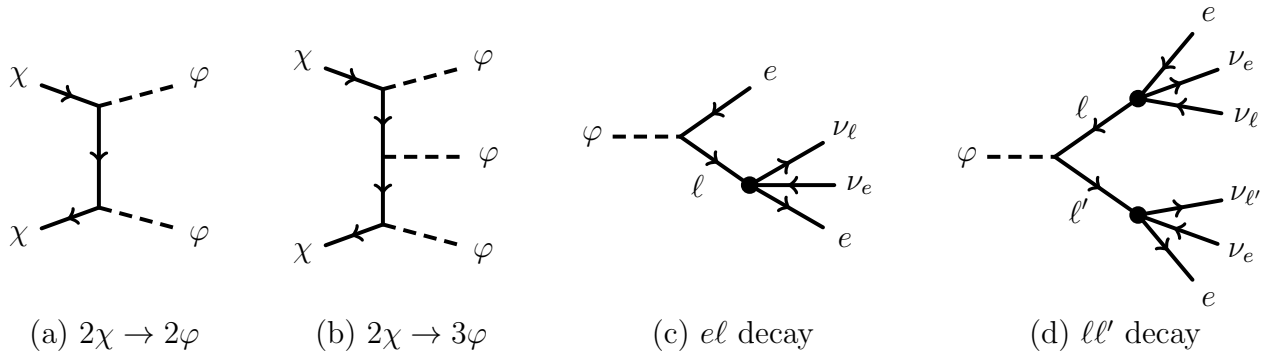
\begin{figure}
\centering
\begin{tabular}{cccc}
	\raisebox{-.5\height}{
	\begin{tikzpicture}[line width=1.5 pt, scale=.9]
	\def\labelscaling{0.4}		% distance of label from endpoint
	\def\labelscalingup{-0.45}	% distance of label from line
	\def\legangle{20}			% angle of legs
	\def\leglength{.9}			% length of legs
	\def\legoffset{0.3}			% horizontal offset for long legs
	\def\labsize{}				% e.g. \large
	\coordinate (v1) at (0, .75); % top vertex
	\coordinate (v2) at (0,-.75); % bot vertex
	\coordinate (leg1a) at ($(v1)-(-\legangle:\leglength)$); 
	\coordinate (leg2a) at ($(v2)-( \legangle:\leglength)$); 
	\coordinate (leg1b) at ($(v1)+( \legangle:\leglength)+(\legoffset,0)$);
	\coordinate (leg2b) at ($(v2)+(-\legangle:\leglength)+(\legoffset,0)$);
%		
%	% label positions
	\coordinate (l1a) at ($(leg1a) - \labelscaling*(-\legangle:\leglength)$); 
	\coordinate (l1b) at ($(leg1b) + \labelscaling*(\legangle:\leglength)$);
	\coordinate (l2a) at ($(leg2a) - \labelscaling*( \legangle:\leglength)$);
	\coordinate (l2b) at ($(leg2b) + \labelscaling*(-\legangle:\leglength)$);
	\draw[fermion, line cap=round] (leg1a)--(v1);
	\draw[fermion, line cap=round] (v1)--(v2);
	\draw[fermion, line cap=round] (v2)--(leg2a);
	\draw[dashed, dash pattern=on 5 off 3] (v1)--(leg1b);
	\draw[dashed, dash pattern=on 5 off 3] (v2)--(leg2b);
%	
%	% draw nodes
	\node at (l1a) {\labsize $\displaystyle \chi$};
	\node at (l1b) {\labsize $\displaystyle \varphi$};
	\node at (l2a) {\labsize $\displaystyle \chi$};
	\node at (l2b) {\labsize $\displaystyle \varphi$};
\end{tikzpicture}

%\begin{tikzpicture}[line width=1.75 pt]
%%
%	\def\labelscaling{0.4}		% distance of label from endpoint
%	\def\labelscalingup{-0.45}	% distance of label from line
%	\def\legangle{20}			% angle of legs
%	\def\leglength{1}			% length of legs
%	\def\legoffset{0.3}			% horizontal offset for long legs
%%
%	\coordinate (v1) at (0, .75); % top vertex
%	\coordinate (v2) at (0,-.75); % bot vertex
%%		
%	\coordinate (leg1a) at ($(v1)-(-\legangle:\leglength)$); 
%	\coordinate (leg2a) at ($(v2)-( \legangle:\leglength)$); 
%%
%	\coordinate (leg1b) at ($(v1)+( \legangle:\leglength)+(\legoffset,0)$);
%	\coordinate (leg2b) at ($(v2)+(-\legangle:\leglength)+(\legoffset,0)$);
%%		
%%	% label positions
%	\coordinate (l1a) at ($(leg1a) - \labelscaling*(-\legangle:\leglength)$); 
%	\coordinate (l1b) at ($(leg1b) + \labelscaling*(\legangle:\leglength)$);
%	\coordinate (l2a) at ($(leg2a) - \labelscaling*( \legangle:\leglength)$);
%	\coordinate (l2b) at ($(leg2b) + \labelscaling*(-\legangle:\leglength)$);
%%
%	\draw[fermion, line cap=round] (leg1a)--(v1);
%	\draw[fermion, line cap=round] (v1)--(v2);
%	\draw[fermion, line cap=round] (v2)--(leg2a);
%%
%	\draw[dashed, dash pattern=on 5 off 3] (v1)--(leg1b);
%	\draw[dashed, dash pattern=on 5 off 3] (v2)--(leg2b);
%%	
%%	% draw nodes
%	\node at (l1a) {\large $\displaystyle \chi$};
%	\node at (l1b) {\large $\displaystyle \varphi$};
%	\node at (l2a) {\large $\displaystyle \chi$};
%	\node at (l2b) {\large $\displaystyle \varphi$};
%%
%\end{tikzpicture}
	}
	&
	\raisebox{-.5\height}{
	\begin{tikzpicture}[line width=1.5 pt, scale=.9]
	\def\labelscaling{0.4}		% distance of label from endpoint
	\def\labelscalingup{-0.45}	% distance of label from line
	\def\legangle{20}			% angle of legs
	\def\leglength{.9}			% length of legs
	\def\legoffset{0.3}			% horizontal offset for long legs
	\def\labsize{}
	\coordinate (v1) at (0, 1); % top vertex
	\coordinate (v2) at (0,-1); % bot vertex
	\coordinate (v3) at (0, 0); % middle vertex
	\coordinate (leg1a) at ($(v1)-(-\legangle:\leglength)$); 
	\coordinate (leg2a) at ($(v2)-( \legangle:\leglength)$); 
	\coordinate (leg1b) at ($(v1)+( \legangle:\leglength)+(\legoffset,0)$);
	\coordinate (leg2b) at ($(v2)+(-\legangle:\leglength)+(\legoffset,0)$);
	\coordinate (leg3b) at ($(v3)+(0:\leglength)+(\legoffset,0)$);
%		
%	% label positions
	\coordinate (l1a) at ($(leg1a) - \labelscaling*(-\legangle:\leglength)$); 
	\coordinate (l1b) at ($(leg1b) + \labelscaling*(\legangle:\leglength)$);
	\coordinate (l2a) at ($(leg2a) - \labelscaling*( \legangle:\leglength)$);
	\coordinate (l2b) at ($(leg2b) + \labelscaling*(-\legangle:\leglength)$);
	\coordinate (l3b) at ($(leg3b) + \labelscaling*(0:\leglength)$);
	\draw[fermion, line cap=round] (leg1a)--(v1);
	\draw[fermion, line cap=round] (v1)--(v3);
	\draw[fermion, line cap=round] (v3)--(v2);
	\draw[fermion, line cap=round] (v2)--(leg2a);
	\draw[dashed, dash pattern=on 5 off 3] (v1)--(leg1b);
	\draw[dashed, dash pattern=on 5 off 3] (v2)--(leg2b);
	\draw[dashed, dash pattern=on 5 off 3] (v3)--(leg3b);
%	
%	% draw nodes
	\node at (l1a) {\labsize $\displaystyle \chi$};
	\node at (l1b) {\labsize $\displaystyle \varphi$};
	\node at (l2a) {\labsize $\displaystyle \chi$};
	\node at (l2b) {\labsize $\displaystyle \varphi$};
	\node at (l3b) {\labsize $\displaystyle \varphi$};
\end{tikzpicture} 
	}
	&
	\raisebox{-.5\height}{
	\begin{tikzpicture}[line width=1.5 pt, scale=.9]
	\def\labelscaling{0.3}	% distance of; label from endpoint
	\def\fangle{35}			% angle for fermions
	\def\fanglem{235}			% angle for fermions
	\def\legin{1}
	\def\leglen{1.25}
	\def\gangle{30}
	\def\blobsize{.1};
	\def\morelabelscale{1.2};
	\def\labsize{}
	\coordinate (e1) at (-\legin, 0); % left vertex
	\coordinate (v1) at (0, 0); % main decay
	\coordinate (f1) at ($(v1) + (\fangle:\leglen)$);
	\coordinate (f2) at ($(v1) + (-\fangle:\leglen)$);
	\coordinate (f3) at ($(f2) + (\gangle:\leglen)$);
	\coordinate (f4) at ($(f2) + (0:\leglen)$);
	\coordinate (f5) at ($(f2) + (-\gangle:\leglen)$);
	\draw[dashed, dash pattern=on 5 off 3] (e1)--(v1);
	\draw[fermion, line cap=round] (f1)--(v1);
	\draw[fermion, line cap=round] (v1)--(f2);
	\draw[fermion, line cap=round] (f2)--(f3);
	\draw[fermion, line cap=round] (f4)--(f2);
	\draw[fermion, line cap=round] (f2)--(f5);
	\draw[fill=black] (f2) circle (\blobsize);
	\coordinate (l1) at ($(e1) - \labelscaling*(\legin,0)$); 
	\coordinate (lf1) at ($(f1) + \labelscaling*(\fangle:\leglen)$); 
	\coordinate (lf3) at ($(f3) + \labelscaling*(\gangle:\leglen)$); 
	\coordinate (lf4) at ($(f4) + \labelscaling*(0:\leglen)$); 
	\coordinate (lf5) at ($(f5) + \labelscaling*(-\gangle:\leglen)$); 
% \coordinate (underleg1) at ($(leg1b) !.5! (v2)$);	
	\coordinate (lm) at ($(v1) !.5! (f2)$); 
	\coordinate (lmm) at ($(lm) + \morelabelscale*\labelscaling*(\fanglem:\leglen)$); 
	\node at (l1) {\labsize $\displaystyle \varphi$};
	\node at (lf1) {\labsize $\displaystyle e$};
	\node at (lf3) {\labsize $\displaystyle \nu_\ell$};
	\node at (lf4) {\labsize $\displaystyle \nu_e$};
	\node at (lf5) {\labsize $\displaystyle e$};
	\node at (lmm) {\labsize $\displaystyle \ell$};
\end{tikzpicture}
	}
	&
	\raisebox{-.5\height}{
	\begin{tikzpicture}[line width=1.5 pt, scale=.9]
	\def\labelscaling{0.3}	% distance of label from endpoint
	\def\fangle{35}			% angle for fermions
	\def\fanglem{235}			% angle for fermions
	\def\legin{1}
	\def\leglen{1.25}
	\def\leglenlong{1.5}
	\def\gangle{30}
	\def\blobsize{.1};
	\def\htangle{10};
	\def\hcangle{-20};
	\def\hbangle{-50}
	\def\morelabelscale{1.2};
	\def\labsize{};
	\coordinate (e1) at (-\legin, 0); % left vertex
	\coordinate (v1) at (0, 0); % main decay
	\coordinate (f1) at ($(v1) + (\fangle:\leglenlong)$);
	\coordinate (f2) at ($(v1) + (-\fangle:\leglenlong)$);
	\coordinate (f3) at ($(f2) + (\htangle:\leglen)$);
	\coordinate (f4) at ($(f2) + (\hcangle:\leglen)$);
	\coordinate (f5) at ($(f2) + (\hbangle:\leglen)$);
	\coordinate (f6) at ($(f1) + (-\htangle:\leglen)$);
	\coordinate (f7) at ($(f1) + (-\hcangle:\leglen)$);
	\coordinate (f8) at ($(f1) + (-\hbangle:\leglen)$);
	\draw[dashed, dash pattern=on 5 off 3] (e1)--(v1);
	\draw[fermion, line cap=round] (f1)--(v1);
	\draw[fermion, line cap=round] (v1)--(f2);
	\draw[fermion, line cap=round] (f2)--(f3);
	\draw[fermion, line cap=round] (f4)--(f2);
	\draw[fermion, line cap=round] (f2)--(f5);
	\draw[fermion, line cap=round] (f6)--(f1);
	\draw[fermion, line cap=round] (f1)--(f7);
	\draw[fermion, line cap=round] (f8)--(f1);
	\draw[fill=black] (f2) circle (\blobsize);
	\draw[fill=black] (f1) circle (\blobsize);
	\coordinate (l1) at ($(e1) - \labelscaling*(\legin,0)$); 
	\coordinate (lf3) at ($(f3) + \labelscaling*(\htangle:\leglen)$); 
	\coordinate (lf4) at ($(f4) + \labelscaling*(\hcangle:\leglen)$); 
	\coordinate (lf5) at ($(f5) + \labelscaling*(\hbangle:\leglen)$); 
	\coordinate (lf6) at ($(f6) + \labelscaling*(-\htangle:\leglen)$); 
	\coordinate (lf7) at ($(f7) + \labelscaling*(-\hcangle:\leglen)$); 
	\coordinate (lf8) at ($(f8) + \labelscaling*(-\hbangle:\leglen)$); 
	\coordinate (lm) at ($(v1) !.5! (f2)$); 
	\coordinate (lmm) at ($(lm) + \morelabelscale*\labelscaling*(\fanglem:\leglen)$); 
	\coordinate (ln) at ($(v1) !.5! (f1)$); 
	\coordinate (lnn) at ($(ln) + \labelscaling*(-\fanglem:\leglen)$); 
	\node at (l1) {\labsize $\displaystyle \varphi$};
%	\node at (lf1) {\labsize $\displaystyle e$};
	\node at (lf3) {\labsize $\displaystyle \nu_{\ell'}$};
	\node at (lf4) {\labsize $\displaystyle \nu_e$};
	\node at (lf5) {\labsize $\displaystyle e$};
	\node at (lf6) {\labsize $\displaystyle \nu_\ell$};
	\node at (lf7) {\labsize $\displaystyle \nu_e$};
	\node at (lf8) {\labsize $\displaystyle e$};
	\node at (lmm) {\labsize $\displaystyle \ell'$};
	\node at (lnn) {\labsize $\displaystyle \ell$};
\end{tikzpicture} 
	}
	\\
	(a) $2\chi \to 2\varphi$ %annihilation 
	& (b) $2\chi \to 3\varphi$ %annihilation
	& (c) $e \ell$ decay  
	& (d) $\ell \ell'$ decay
\end{tabular}
	\caption{Diagrams showing annihilation to mediators and possible off-diagonal mediator decay modes.
	The dot is a Fermi vertex from an off-shell $W$ for the subsequent decay of heavy leptons. $\tau$ final states also carry a large hadronic branching ratio.
	}
	\label{fig:off:shell:heavy:lepton:diagrams}
\end{figure}

We present our model in this section; the interactions are summarized in Fig.~\ref{fig:off:shell:heavy:lepton:diagrams}.

\subsection{Dark Sector Interactions}

Fermionic dark matter $\chi$ is proposed to interact with a spin-0 mediator, $\varphi$, through the interactions,
\begin{align}
	\mathcal L_\varphi\chi &= 
	\frac 12 y_S\eta  \varphi \bar\chi \chi 
	+ \frac i2 y_P \eta \varphi \bar\chi \gamma^5\chi  
	&
	\eta &= 1\,\left(2\right) 
	\;\, \text{for Majorana (Dirac) $\chi$} \ .
	\label{eq:L:phi:chi}
\end{align}
If $\varphi$ is complex, one must also include $\mathcal L_{\varphi\chi}^\dag$. 
Unlike the case of a vector mediator for which there is no coupling to Majorana dark matter, the spin-0 mediator can couple to either Dirac or Majorana fermions. 
In writing these interactions, we assume that the fermion mass is manifestly real so that the pseudoscalar interaction is physical. See, for example, Ref.~\cite{Fedderke:2014wda} for a demonstration of how a complex fermion mass term---say, if $\langle \varphi\rangle$ and $y_P \neq 0$---would shift the pseudoscalar interaction upon performing a chiral rotation to make the mass term real. For the remainder of this manuscript we assume that $\langle \varphi \rangle = 0$.

In the secluded dark matter scenario where annihilation goes into on-shell mediators, the dark matter--mediator couplings of (\ref{eq:L:phi:chi}) control the annihilation rate independently of the mediator--Standard Model couplings. A useful benchmark is the annihilation cross-section required for $\chi$ to reproduce the observed dark matter density assuming that it was initially in thermal equilibrium with the Standard Model~\cite{Steigman:2012nb}, 
\begin{align}
\langle \sigma v \rangle_\text{rel.} = 2.2\eta \times 10^{-26} \text{cm}^3/\text{s}	\ .
\label{eq:relic:abundance:xsec}
\end{align}
Below we present targets for the dark matter--mediator couplings implicitly as a ratio of the required present day annihilation cross-section, $\langle\sigma v\rangle$ set by the Fermi $\gamma$-ray excess, to the thermal relic cross-section, $\langle\sigma v\rangle_\text{rel.}$. In doing so, we cancel all dependence on $\eta$ in our results.
The extent to which $\langle\sigma v\rangle$ is compatible with $\langle\sigma v\rangle_\text{rel.}$ is a useful metric of attractiveness for our models. Compatibility is simplest when annihilation is dominantly $s$-wave since higher partial waves are highly velocity suppressed in the present day. In this case, the dark matter and mediator masses are in the range 
\begin{align}
	m_\chi &\sim \mathcal O(10-100\text{ GeV})
	&
	m_{\ell_i} + m_{\ell_j} < m_\varphi < m_\chi~\left(\frac23 m_\chi\right) 
	\label{eq:mass:ranges}
\end{align}
where the value in parenthesis corresponds to the $2\to 3$ annihilation.

The $s$-wave annihilation of fermionic dark matter into on-shell spin-0 mediators was recently studied in Refs.~\cite{Abdullah:2014lla, Rajaraman:2015xka, Cline:2015qha}. There are three scenarios which give different $s$-wave annihilation modes depending on the parity of $\varphi$:
\begin{enumerate}
	\item If $\varphi$ does not have a well defined dark sector parity---that is, both $y_S$ and $y_P \neq 0$---then the leading $s$-wave annihilation mode is $\chi\bar\chi \to \varphi\varphi^*$. This holds for both real and complex $\varphi$. This is shown in Fig.~\ref{fig:off:shell:heavy:lepton:diagrams}(a).
	\item If $\varphi$ is a parity-odd pseudoscalar ($y_S = 0$), then the leading $s$-wave annihilation mode is $\chi\bar\chi \to 3\varphi$. This is shown in Fig.~\ref{fig:off:shell:heavy:lepton:diagrams}(b).
	\item If $\varphi$ is a parity even scalar ($y_P = 0$), then the leading $s$-wave annihilation mode is $\chi\bar\chi \to 4\varphi$. 
\end{enumerate}
We assume that the mediator mass $m_\varphi$ is sufficiently light such that these annihilation modes are accessible.
In the present work we focus on the first two of these scenarios since they offer viable candidates for the Fermi $\gamma$-ray excess that are compatible with a thermal relic. 
If the annihilation results in additional final state mediators, a heavier dark matter mass is required in order to account for the same energy spectra. This implies a smaller galactic halo dark matter number density which in turn forces the annihilation rate to be larger and typically in further tension with a standard thermal relic.
While we focus on the case when the present day annihilation rate is compatible with thermal freeze out, we remark that one could relax this requirement---as in Kaplinghat et al.~\cite{Kaplinghat:2015gha}---and assume that the dark sector and Standard Model were reheated to different initial temperatures in the early universe. 

Typically one assumes that the mediator decays into Standard Model states: if the mediator is stable, then it is a dark matter component that couples directly to the Standard Model. If, on the other hand, it decays into lighter hidden sector states, then those states are constrained by cosmological limits on the number of light degrees of freedom.

From the point of view of softening the positron spectrum, one may wish to consider mediator masses lighter than one of the leptons to which it couples. In this case, the heavier leptons in Figs.~\ref{fig:off:shell:heavy:lepton:diagrams}(c,d) are off shell. This forces the $\varphi$ decay to be multi-body and further softens the spectrum. We do not consider this possibility because the constraints from two-body decays of heavy leptons~\cite{Albrecht:1995ht,Bayes:2014lxz}, typically imply that the mediator is too long-lived and would be ruled out by cosmological bounds.

\subsection{Standard Model Interactions}

Both the dark matter and mediator are taken to be Standard Model singlets.
We assume that at low energies, the mediator communicates to the Standard Model only through charged lepton interactions that are chiral and flavor violating,
\begin{align}
	\mathcal L_{\varphi \, \text{SM}} &=  
	g_{ij}
	\varphi \bar \ell_i
	P_{L}
	 \ell_j 
	+ g^*_{ji} \varphi^* \bar\ell_j P_R \ell_i
	 \ ,
	\label{eq:L:phi:SM}
\end{align}
where $P_{L,R} = \frac 12(1\mp\gamma^5)$ are chiral projection operators.
We restrict to the case where $g_{ij}$ is non-zero for only a single pair of distinct flavors, $i\neq j$.
For a complex mediator, we choose as a convention,
that $\varphi$ rather than $\varphi^*$ interacts with a left-handed projection operator. The dominant decay modes of this mediator are shown in Fig.~\ref{fig:off:shell:heavy:lepton:diagrams}(c,d). 
We demonstrate below that the chiral flavor violating structure of this interaction softens constraints and better fits the Fermi $\gamma$-ray excess  than the Kaplinghat et al.\ model~\cite{Kaplinghat:2015aga}.
This low-energy model is invariant under a global $L_i - L_j$ symmetry which prohibits most charged lepton-flavor violating processes up to corrections from the $W$ interactions and neutrino masses. These are discussed in Section~\ref{eq:constraints}.

\subsection{Comments on Ultraviolet Realizations}

While it is beyond the scope of this phenomenological study, we briefly comment on complete flavor models that may produce this scenario.
At the electroweak scale, the interactions in (\ref{eq:L:phi:SM}) can be generated by the gauge-invariant, higher-dimensional operator
\begin{align}
	\mathcal L_{\varphi\text{SM}}^{(\text{EW})} 
	& = 
	\frac{c_{ij}}{\Lambda} \varphi H \bar L_i E_j
	+ \text{h.c.} \ ,
	\label{eq:L:phi:SM:EW}
\end{align}
so that one may identify $g_{ij} = c_{ij}v/\sqrt{2}\Lambda$ and $v$ is the Higgs vacuum expectation value. 
As an alternative, one could assume (\ref{eq:L:phi:SM}) is generated by $R$-parity violating type superpotential interaction $\lambda_{ijk}L^iL^k\bar E^k$ in a supersymmetric extension of the Standard Model, where $i,j,k$ are distinct lepton flavors.
This superpotential generates a chiral, flavor-changing coupling mediated by sneutrinos~\cite{Halprin:1993zv, Mohapatra:1991ij}.

The electroweak-scale effective operator (\ref{eq:L:phi:SM:EW}) can, in turn, be generated by renormalizable interactions with respect to heavy degrees of freedom. This may be realizable within the Froggatt--Nielsen mechanism, in which a pattern of low-energy flavor-dependent couplings is generated by integrating out heavy degrees of freedom that are charged under the broken flavor symmetries~\cite{Froggatt:1978nt}. This mechanism is typically used to explain the pattern of Standard Model Yukawa couplings, but is straightforwardly extended to additionally generate the coupling in (\ref{eq:L:phi:SM:EW}).
We reserve the model building aspects for future work,
but remark that one can build such realizations in which
(\ref{eq:L:phi:SM:EW}) has only a single dominant
entry which is off-diagonal, with all other entries suppressed.
In this case, the scale $\Lambda$ of the effective operator
can readily be associated with the flavor symmetry
breaking scale, and need not be much higher
than the electroweak scale, 
so that the renormalization group corrections to the pattern are very small.
While neutrinos may modify the flavor structure of the mediator--Standard Model interactions at two-loop order; we ignore these as they are typically several orders of magnitude smaller than $g_{ij}$. 
Supersymmetric flavor models may be a promising direction for such model-building~\cite{Leurer:1992wg, Nir:1993mx, Leurer:1993gy}, especially those in which the flavor symmetry is broken at a higher scale than supersymmetry~\cite{Shadmi:2011hs,Abdullah:2012tq,Galon:2013jba}.
In this case supersymmetry protects the flavor structure from renormalization group effects down to the supersymmetry breaking scale, and possibly down to the scale of slepton masses.

\subsection{Self-Interacting Dark Matter}

The scalar coupling $y_S$ in (\ref{eq:L:phi:chi}) generates a long-range Yukawa potential between dark matter particles. This is the key ingredient for how dark matter self-interactions address small scale structure anomalies~\cite{Tulin:2013teo}. 
The lightness of the mediator introduces a velocity-dependence on the self-scattering cross-section; this affects the dark matter halo profile on the scales of dwarf galaxies, while remaining consistent with constraints from galaxy cluster mergers~\cite{Harvey:2015hha}.  
In the model by Kaplinghat, Linden, and Yu, a vector mediator, $V$, produces the Fermi $\gamma$-ray excess by the inverse Compton scattering of $\chi\bar\chi \to VV \to 4e$~\cite{Kaplinghat:2015gha}. The vector mass, $m_V \sim \mathcal O(10-100~\text{MeV})$, and coupling to dark matter, with a transfer cross-section of $\sigma_T\sim \mathcal O(0.5-50~\text{cm}^2/\text{s})$, were found to be of the correct size to realize this self-interacting dark matter target region. This came at the cost of some tension with the thermal relic cross-section, $\langle\sigma v \rangle_\text{rel.}$, so that they invoke a different dark sector temperature~\cite{Feng:2008mu}.

In the scalar models here, only the parity-even dark sector interaction in (\ref{eq:L:phi:chi}) mediates a Yukawa potential. The parity-odd interaction mediates a spin dependent potential that scales as $e^{-m_\varphi r}/r^3$~\cite{Bellazzini:2013foa}; this is not expected to have a significant effect on astrophysical dynamics.
We thus observe that the scenario with a pure pseudoscalar mediator ($y_S=0$) does not realize the self-interacting dark matter target region.
On the other hand, in the scenario where $\varphi$ has mixed parity, $s$-wave $\chi\bar\chi \to \varphi \varphi^{(*)}$ annihilation depends on both $y_S$ and $y_P$. This introduces some freedom to choose $y_S$ to realize a large self-interaction cross-section and then separately choose $y_P$ to select the annihilation cross-section, $\langle\sigma v\rangle$. 
Our models also differ from Kaplinghat et al.\ because the minimum mediator mass scale is set by the heavier lepton to which the mediator couples. Thus the lightest mediator mass we consider is $m_\varphi \sim m_\mu = 106~\text{MeV}$ which is accessible for mediators with $\mu e$ couplings. Observe that this mass is near the heavy limit of mediator masses that are compatible with solving small scale structure anomalies~\cite{Tulin:2013teo}. We then expect that the case where the mediator couples to a $\tau$ are typically incompatible with the self-interacting dark matter target region.

While a detailed study of the dark matter self-interactions in this model is beyond the scope of this paper, the benchmark results in Ref.~\cite{Tulin:2013teo} already demonstrate the key properties. We specifically note that the dark matter and mediator masses considered here populate the numerically difficult resonant regime where consistency with the self-interacting dark matter target region is plausible but very sensitive to the precise values of $m_{\chi,\varphi}$. For this reason, in this manuscript we focus on the compatibility of the Fermi $\gamma$-ray excess in our scenario with the thermal relic cross-section without invoking a different dark sector temperature.  We leave the details of the dark sector self-interactions---which we emphasize are automatic in our constructions---for separate work.

\section{The Fermi-LAT $\gamma$-Ray Excess}
\label{sec:fermi}

\subsection{Photons from Leptons}
\label{sec:photons:from:leptons}

The spectrum of photons to be identified with the Fermi $\gamma$-ray excess originate from two sources:
\begin{enumerate}
	\item Prompt photons from the final state leptons or
	\item Up-scattered starlight from the inverse Compton scattering (ICS) of $e^\pm$ produced through the $\varphi$ decay.
\end{enumerate}
This is in contrast to models where dark matter annihilates predominantly into quarks or gluons. In that case the photons are a result of $\pi^0\to \gamma\gamma$ decays from the showering and hadronization of the final state partons. Since the $\tau$ has a large hadronic decay width ($\sim65\%$), its spectrum of prompt photons is similar to that of quarks and gluons.  
In contrast,
electrons and muons (and leptonically decaying taus) typically produce a smaller flux of prompt photons, but can yield a large number of up-scattered photons from inverse Compton scattering in a stellar radiation field. 
The scattered photon energy, $E'_\gamma$ can be approximated
in terms of the incoming photon energy, $E_\gamma$ 
and the scattering electron energy, $E_e$
\begin{align}
	E'_\gamma\approx \left(\frac{E_e}{m_e}\right)^2 E_\gamma \ .
	\label{eq:starlight:energy:upscatter}
\end{align}
In the $\chi\bar\chi \to VV\to 4e$ model of Kaplinghat et al., $m_\chi \sim 10~\text{GeV}$ so that $(E_e/m_e)\sim 10^4$ and $\mathcal O(10~\text{eV})$ starlight is then up-scattered to $\mathcal O(\text{GeV})$, corresponding to the characteristic scale of the Fermi-LAT $\gamma$-ray excess.
In contrast, the $e^\pm$ energy spectrum resulting from the 
scalar mediator decays $\varphi\to \bar \ell_i\ell_j  (i\ne j)$
and subsequent decays to electrons 
is softer as part of the energy is deposited in neutrinos.
The average $e^\pm$ energies in the $e\mu,~e\tau,~\mu\tau$
scenarios in the limit $m_{\ell} \ll m_\varphi$ are suppressed by $\sim \frac 23,~\frac{11}{18},~\frac{5}{18}$ with respect to the $\varphi\to e^+e^-$ scenario, where we only account for leptonic $\tau$ decays.
We therefore expect that the average ICS photon energy
to be approximately an order of magnitude 
softer in the flavor-violating case.

In order to properly estimate the photon spectrum, we use the \textit{Mathematica} package \texttt{PPPC}~\cite{Cirelli:2010xx, Buch:2015iya}.
As an input, \texttt{PPPC} requires the flavor dependent 
energy spectrum distributions of the
leptonic annihilation products, 
\begin{align}
	\frac{dN_{\ell_j}}{dE_{\ell_j}}
	=
	\int dE_\varphi\;
	\frac{dN_\varphi}{dE_\varphi}\,
%	\left(
		\frac{d N^{\varphi}_{\ell_j}(E_\varphi)}{dE_{\ell_j}}
%	\right)_\text{lab} 
	\ ,
	\label{eq:dNl:dEl}
\end{align}
where $dN_\varphi/dE_\varphi$ is the spectrum of mediators and $d N^{\varphi}_{\ell_j}(E_\varphi)/dE_{\ell_j}$ is the spectrum of $j$-type leptons produced in the decay of a mediator with energy $E_\varphi$. 
For annihilation into two mediators, the $\varphi$ energy spectrum is monochromatic,
$dN_\varphi/dE_\varphi = \delta(E_\varphi - m_\chi)$,
so that the boost of lepton energies from the $\varphi$
rest-frame is 
\begin{align}
	E_\ell = \gamma E^0_\ell + \sqrt{\gamma^2-1} |\mathbf{p}^0_\ell| \cos \theta \ ,
\end{align}
where $E^0_\ell$ and $\mathbf{p}^0_\ell$ are the lepton energy and three-momentum in the $\varphi$ frame, and $\gamma = E_\varphi/m_\varphi$ is the boost to the lab frame. The lepton energy distribution is then box-shaped because the cosine of the azimuthal angle $\cos \theta$ is uniformly distributed over its range.
For annihilation into three mediators, these distributions are implemented following the discussion in Ref.~\cite{Rajaraman:2015xka}; see \cite{Mardon:2009rc, Elor:2015tva} for related discussions on cascade decays from mediators.

The secondary photon spectrum is,
\begin{align}
	\frac{dN_\gamma}{dE_\gamma}
	&=
	\sum_j 
	\int dE_{\ell_j} \;
	\frac{dN_{\ell_j}}{dE_{\ell_j}} \,
	\frac{dN_\gamma^\ell(E_{\ell_j})}{dE_\gamma} \ ,
\end{align}
where $dN_\gamma^\ell(E_\ell)/dE_\gamma$ is the spectrum of photons produced from a lepton $\ell$ with energy $E_\ell$. These are extracted from Pythia~\cite{Sjostrand:2006za} and encoded in \texttt{PPPC}.

\texttt{PPPC} decays muons and taus
and calculates the total $e^\pm$ energy spectrum
at the galactic center region.
In turn, this spectrum is used as an initial condition
in the calculation of the differential $e^\pm$
flux $\frac{d\phi_{e^\pm}}{dE_e}\left(E_e, \vec x\right)$
which determines the inverse Compton scattering spectrum of $\gamma$-rays~\cite{Lacroix:2014eea}.
In calculating the ICS spectrum of photons, we use the \textsc{med} set of diffusion parameters as described in \cite{Delahaye:2007fr} and a Navarro-Frenk-White (NFW) halo profile for the dark matter distribution with inner profile slope of $\gamma_{\text{\tiny{NFW}}}=1.0$ and a local dark matter density of $\rho_{\odot}=0.4 \text{~GeV cm}^{-3}$~\cite{Salucci:2010qr}. 
If one uses a contracted NFW profile, as suggested in Refs.~\cite{Abazajian:2014fta, Daylan:2014rsa, Calore:2014xka} and used by Kaplinghat et al., the required cross-section is reduced by a factor of $\sim 3$ to match the $\gamma$-ray excess intensity.
Varying the diffusion parameters across the range of uncertainties as described in \cite{Delahaye:2007fr} does not significantly affect the resultant ICS spectrum.
However, their theoretical uncertainties are set by modeling the propagation from across much longer ranges than our region of interest.
The actual range of uncertainties on diffusion parameters for galactic center are unknown and may well be much larger.

\subsection{Fit to Fermi $\gamma$-ray Excess}

%\clearpage
\begin{figure}[]
\centering
\begin{subfigure}{0.45\textwidth}
\includegraphics[width=\textwidth]{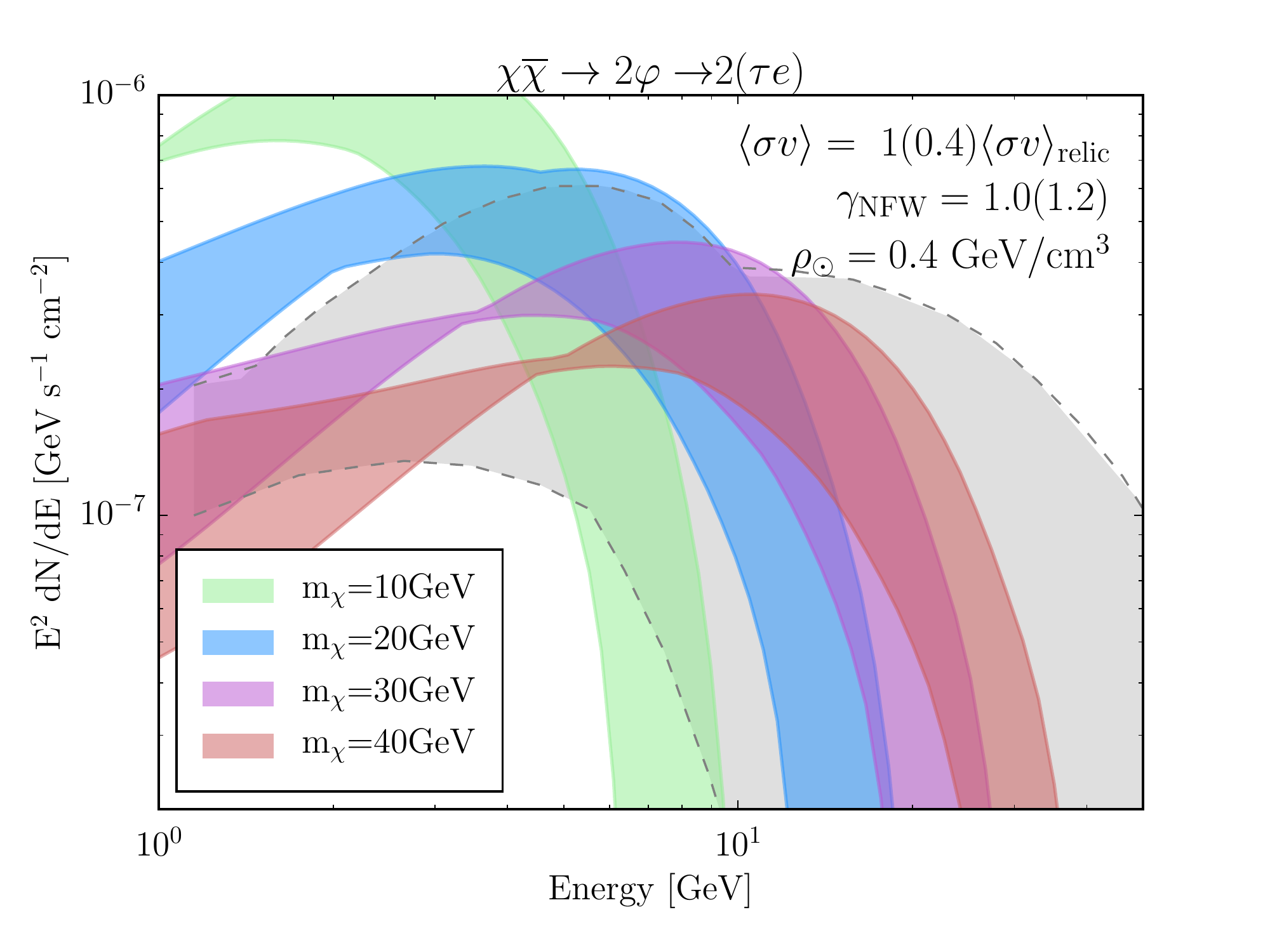}
%\caption{$ $}
\label{fig:}
\end{subfigure}
\begin{subfigure}{0.45\textwidth}
\includegraphics[width=\textwidth]{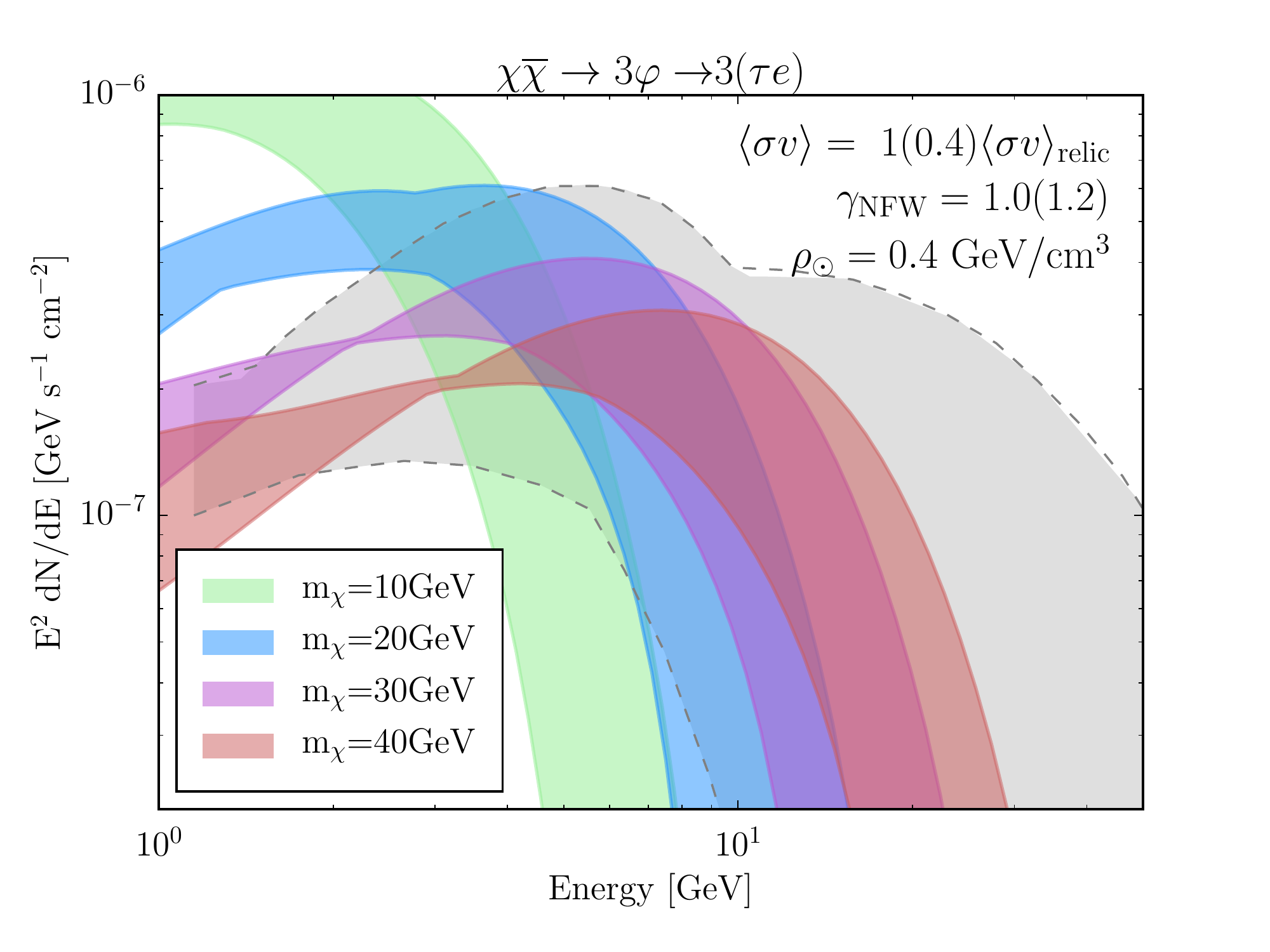}
%\caption{$ $}
\label{fig:}
\end{subfigure}
\begin{subfigure}{0.45\textwidth}
\includegraphics[width=\textwidth]{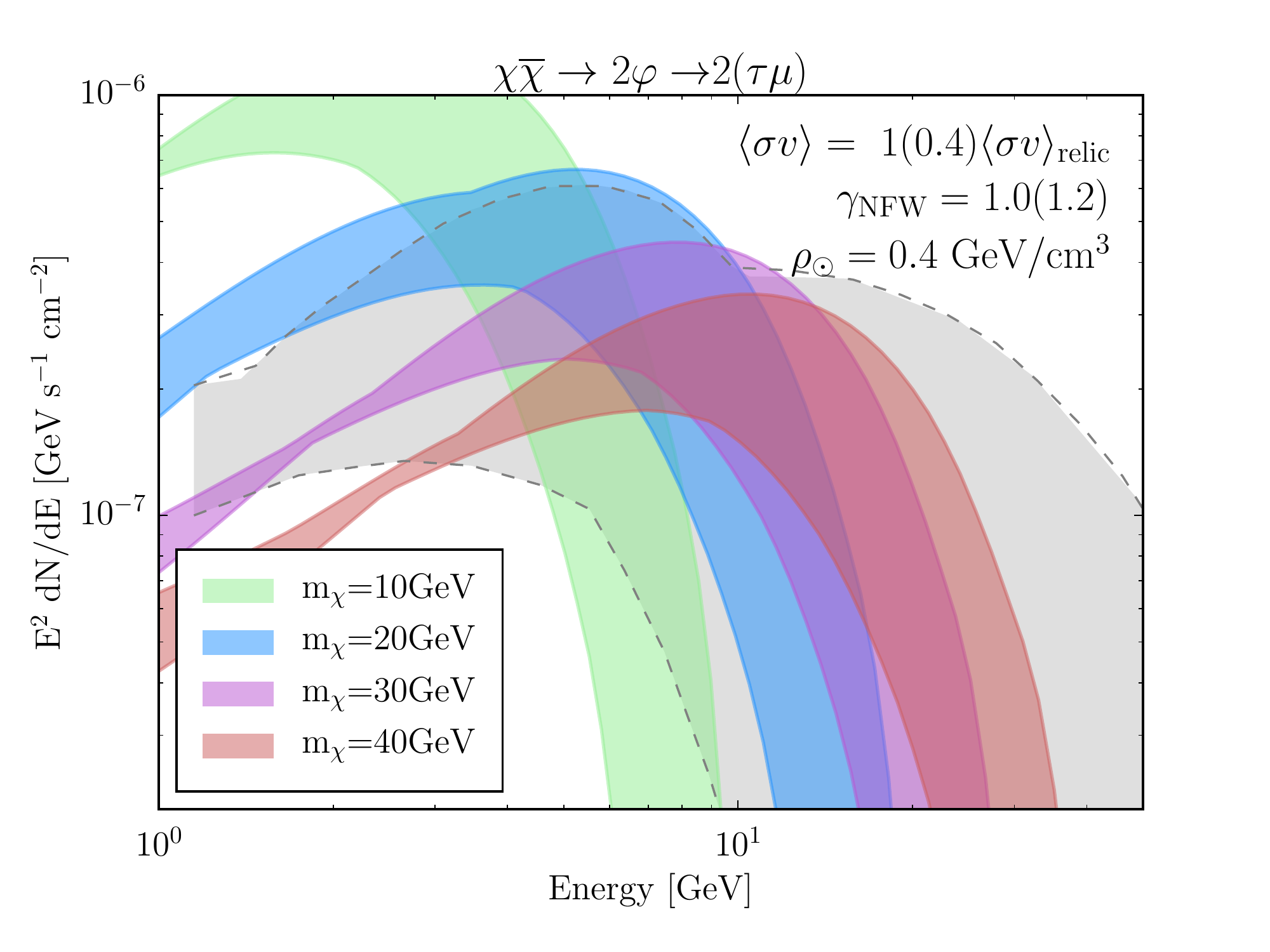}
%\caption{$ $}
\label{fig:}
\end{subfigure}
\begin{subfigure}{0.45\textwidth}
\includegraphics[width=\textwidth]{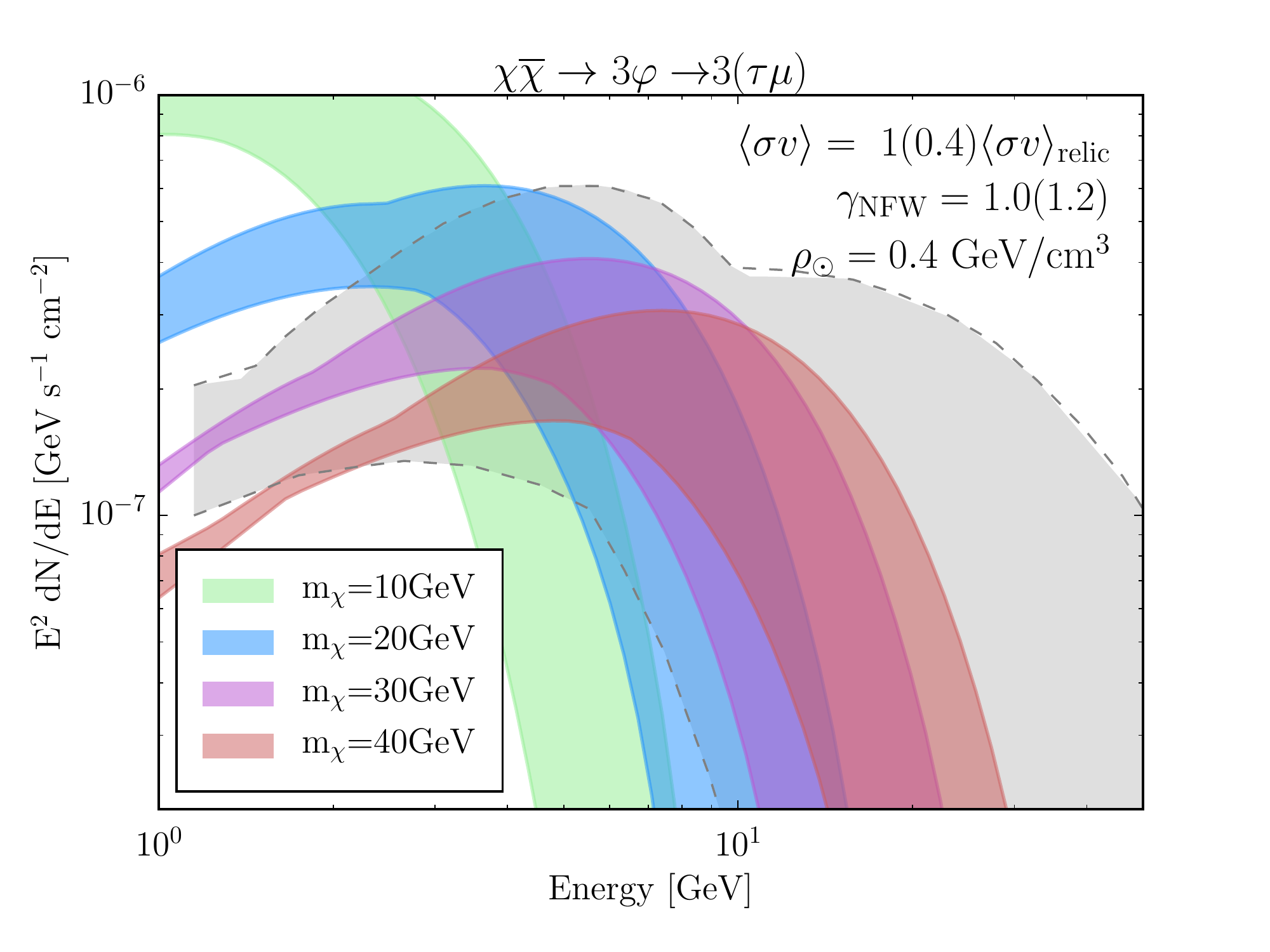}
%\caption{$ $}
\label{fig:}
\end{subfigure}
\begin{subfigure}{0.45\textwidth}
\includegraphics[width=\textwidth]{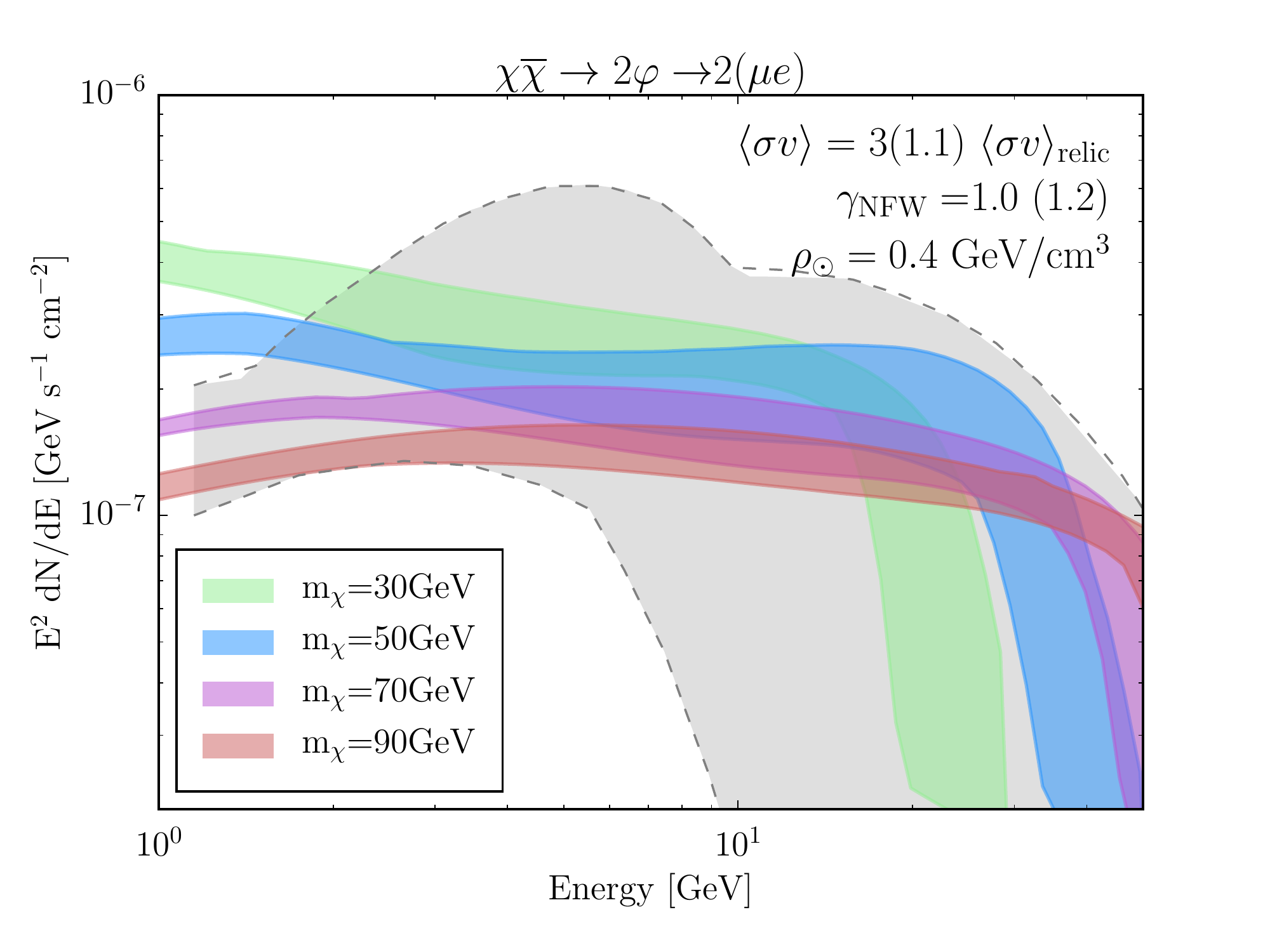}
%\caption{$ $}
\label{fig:}
\end{subfigure}
\begin{subfigure}{0.45\textwidth}
\includegraphics[width=\textwidth]{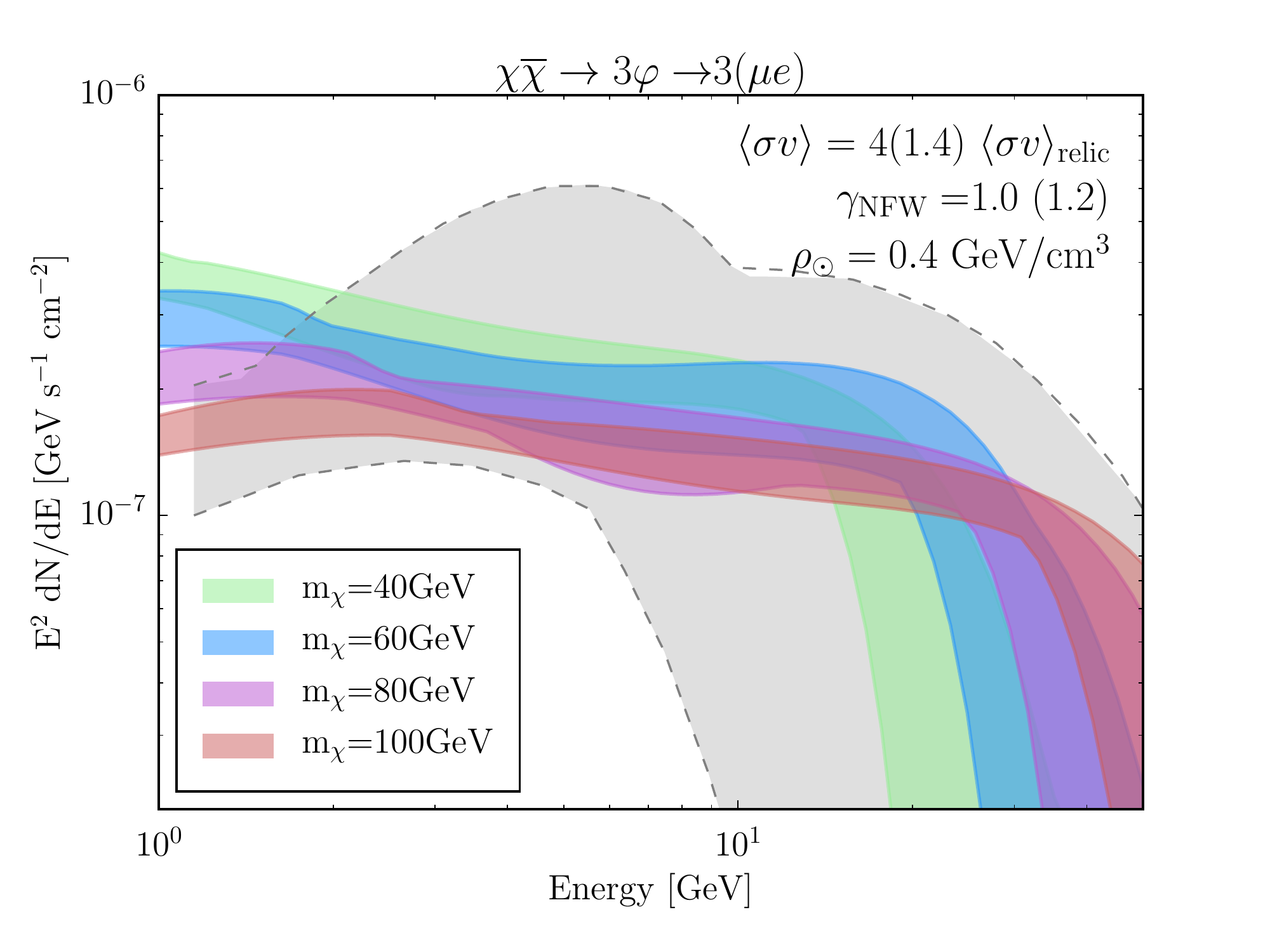}

%\caption{$ $}
\label{fig:}
\end{subfigure}
\caption{
The predicted combined prompt and ICS gamma-ray spectra for each scenario: $\chi\bar\chi\to 2\varphi$ (left) and $\chi\bar\chi\to 3\varphi$ followed by the lepton-flavor violating decays indicated in each panel.
 The grey shaded region represents the Fermi collaboration's $\gamma$-ray excess spectrum bounded by its estimated systematic error when fit as a parameterized form to the entire energy range of the data. Each color-coded band corresponds to a set of $\{m_\chi,~m_\varphi \}$ with $m_\varphi$ varying in the range
$[m_{\ell_{\text{heavy}}},~m_\chi]$ (left) and $[m_{\ell_{\text{heavy}}},~\frac23 m_\chi]$
(right).
The spectra are calculated assuming a halo profile slope of $\gamma_\text{\tiny NFW}=1.0$ and the annihilation cross sections indicated in each figure. For a steeper halo profile of $\gamma_\text{\tiny NFW}=1.2$, cross sections are a factor of $\sim 3$ smaller.% In this scenario the spectral shape may differ slightly due to the non-uniformity of the radiation field in the region of interest; we expect it still remain well within the systematic uncertainties.
}
\label{fig:GCEspectra_parameterized}
\end{figure}
%

%\clearpage
\begin{figure}[]%[h!]
\centering
\begin{subfigure}{0.45\textwidth}
\includegraphics[width=\textwidth]{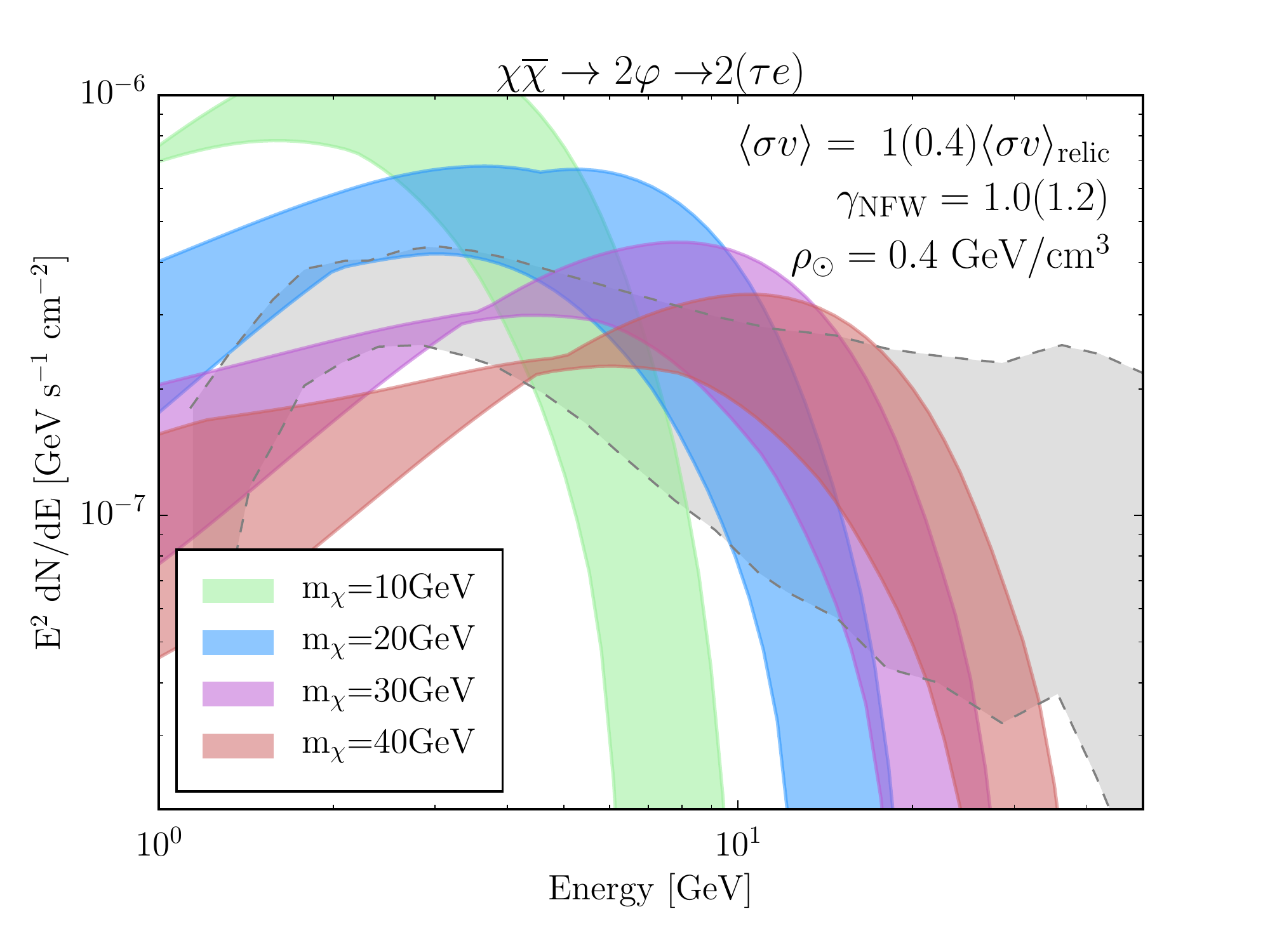}
%\caption{$ $}
\label{fig:}
\end{subfigure}
\begin{subfigure}{0.45\textwidth}
\includegraphics[width=\textwidth]{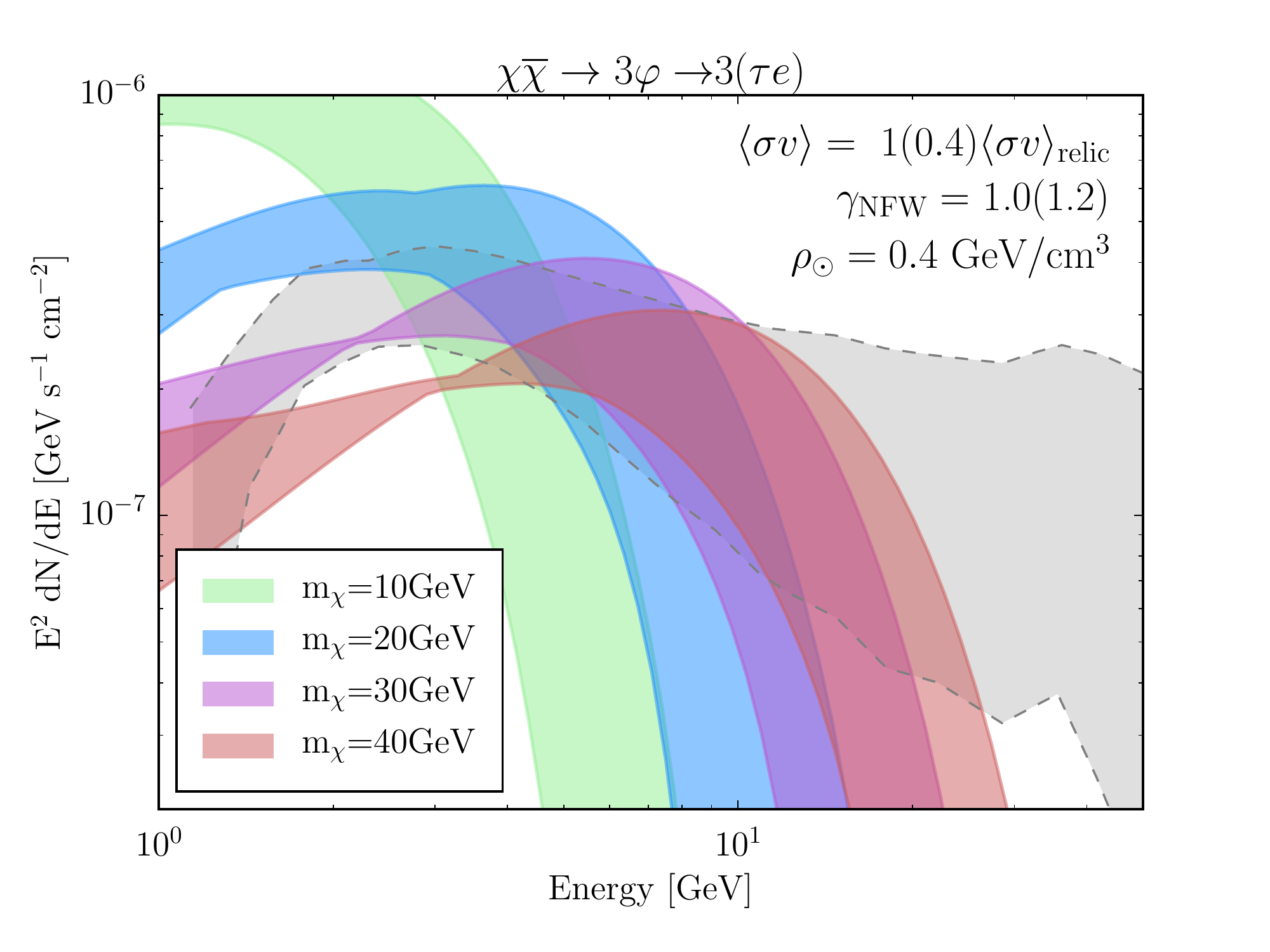}
%\caption{$ $}
\label{fig:}
\end{subfigure}
\begin{subfigure}{0.45\textwidth}
\includegraphics[width=\textwidth]{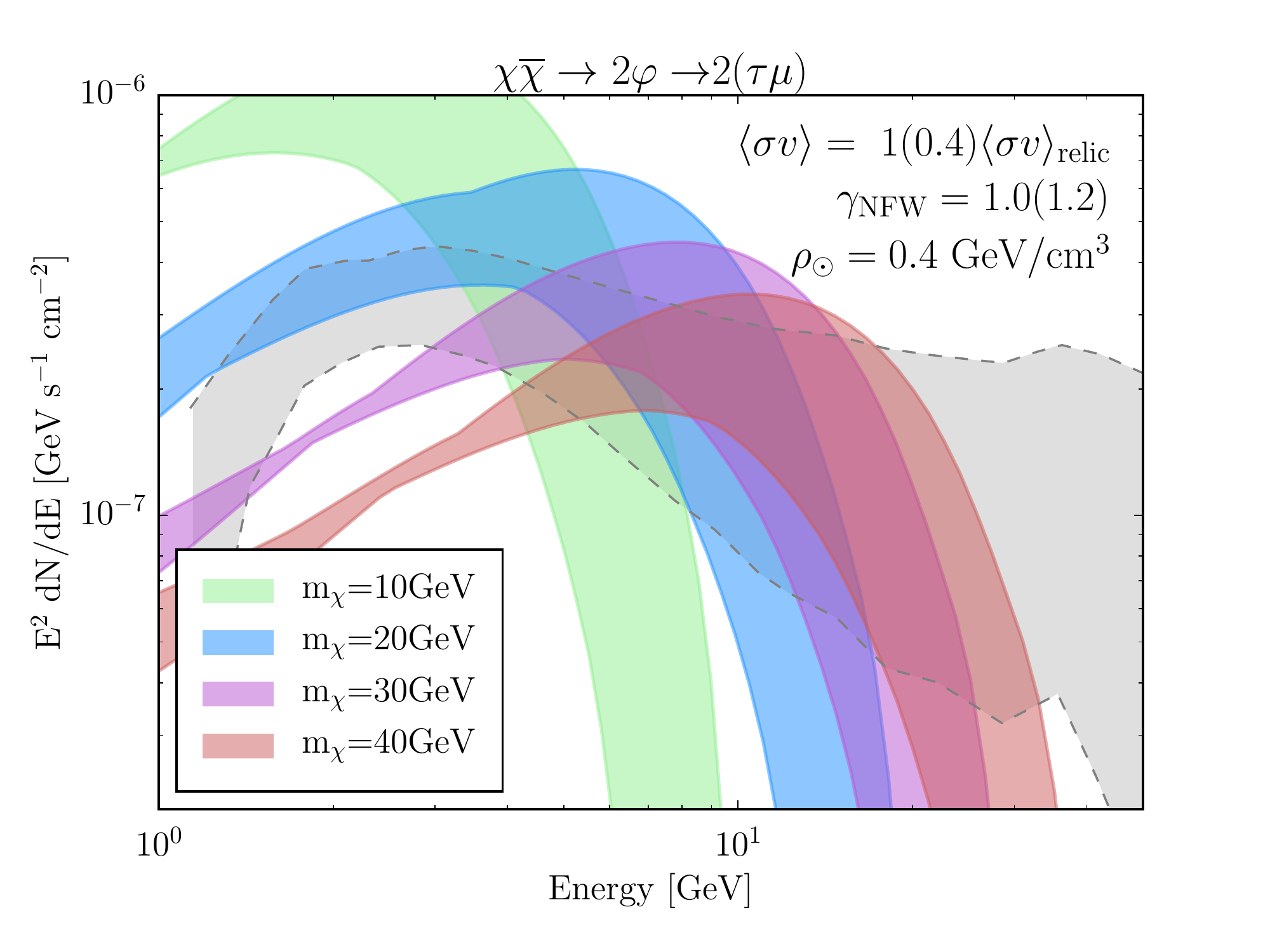}
%\caption{$ $}
\label{fig:}
\end{subfigure}
\begin{subfigure}{0.45\textwidth}
\includegraphics[width=\textwidth]{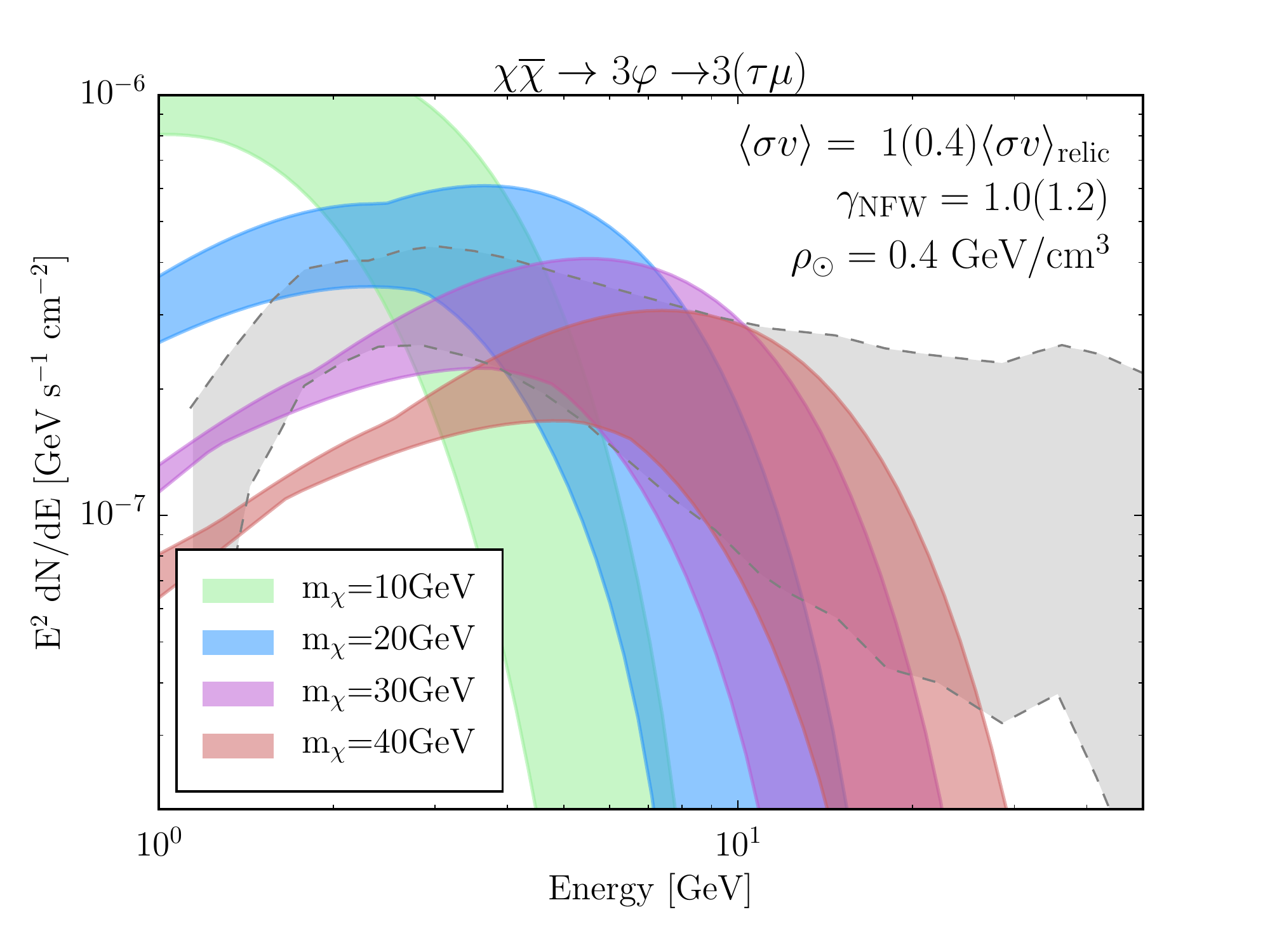}
%\caption{$ $}
\label{fig:}
\end{subfigure}
\begin{subfigure}{0.45\textwidth}
\includegraphics[width=\textwidth]{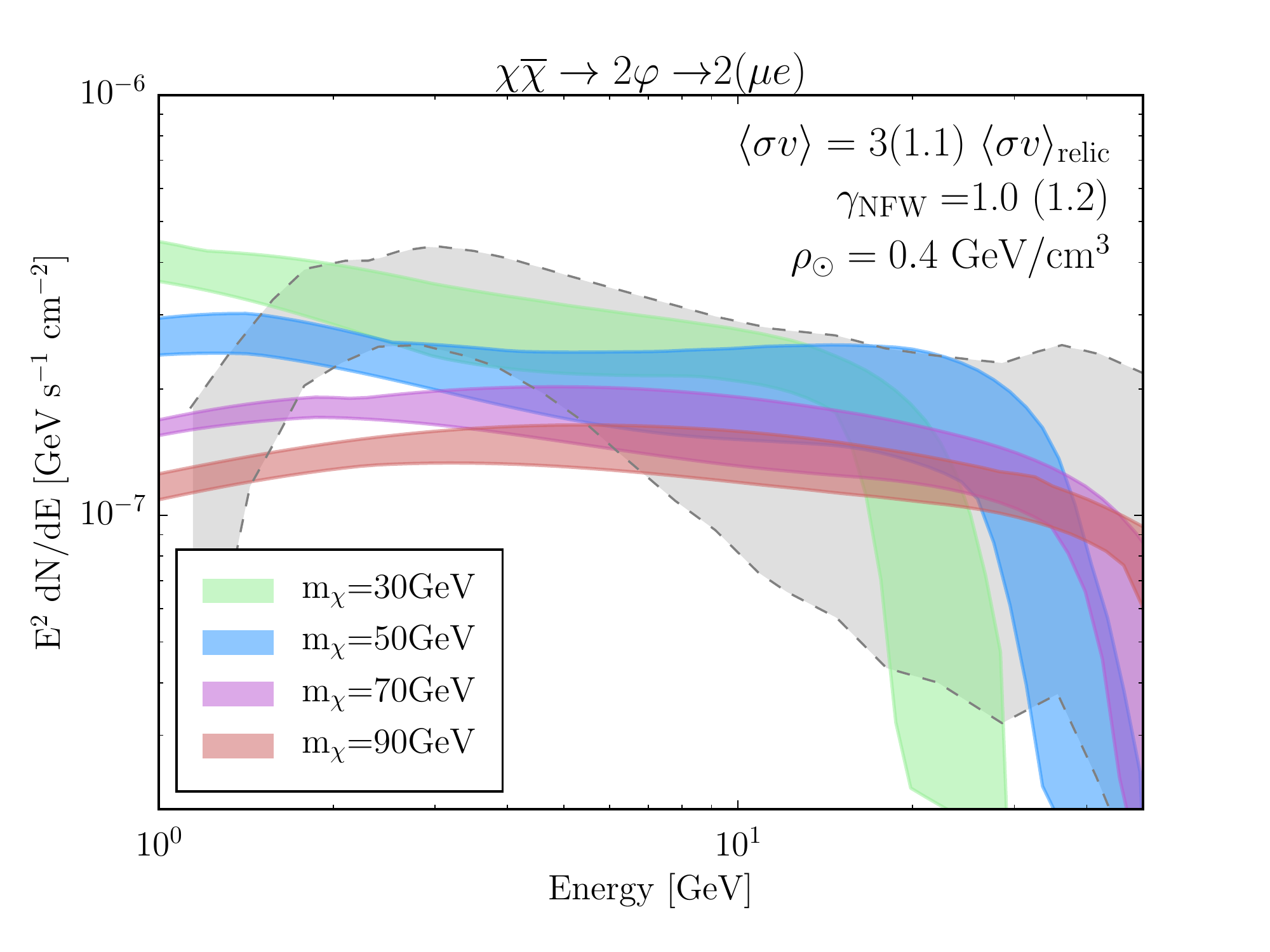}
%\caption{$ $}
\label{fig:}
\end{subfigure}
\begin{subfigure}{0.45\textwidth}
\includegraphics[width=\textwidth]{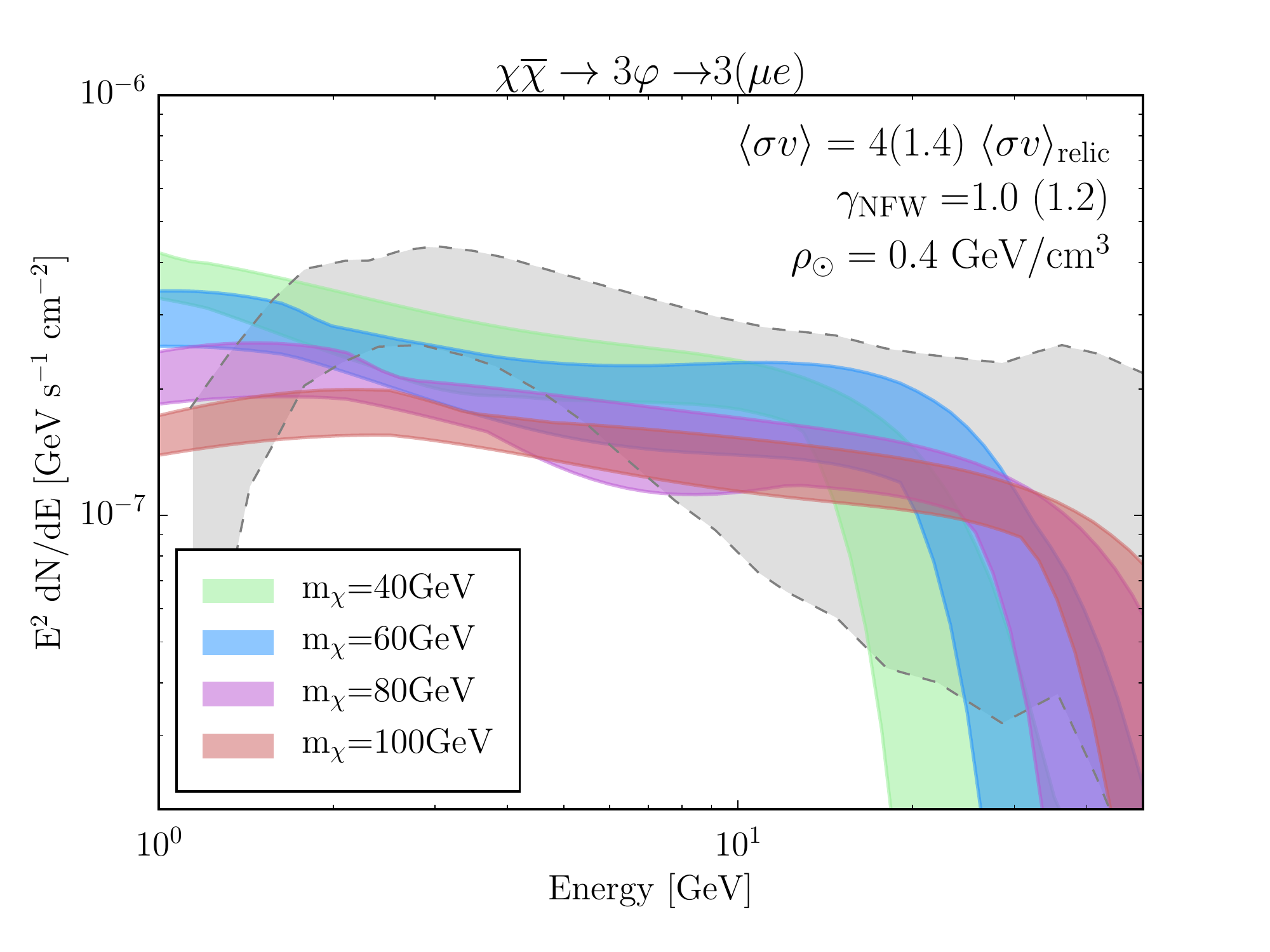}
%\caption{$ $}
\label{fig:}
\end{subfigure}
\caption{
Same as Fig. \ref{fig:GCEspectra_parameterized}, but now the grey shaded region represents the $\gamma$-ray excess spectrum bounded by its estimated systematic error when fit in independent energy bins, as reported by the Fermi collaboration. Our predicted combined prompt and ICS gamma-ray spectra for each scenario: $2\to2$ (left) and $2\to3$ (right) $\chi\chi$ annihilations to $\varphi$s, followed by one of the following decays:
$\varphi\to \tau e$ (up), $\tau\mu$ (middle), and $\mu e$ (bottom). Each color-coded band corresponds to a set of $\{m_\chi,~m_\varphi \}$ with $m_\phi$ varying in the range
$[m_{\ell_{\text{heavy}}},~m_\chi]$ (left) and $[m_{\ell_{\text{heavy}}},~\frac23 m_\chi]$ (right).}
\label{fig:GCEspectra_sys_eng_bin}
\end{figure}

Figs.~\ref{fig:GCEspectra_parameterized} and~\ref{fig:GCEspectra_sys_eng_bin} show the photon spectrum prediction for the 15$^{\circ}\times 15^{\circ}$ region of the sky centered at galactic coordinates $(l, b)=(0,0)$.
We considered each lepton flavor model and each dark matter annihilation mode separately.
We tested several $m_\chi$ benchmarks, each of which is plotted in a different color. In each benchmark, the range of $m_\varphi$ masses considered is accounted for by the thickness of each plotted color. The color edges interpolate the range $m_\varphi \in [m_\ell^{\text{heavy}}, m_\chi~(\frac23 m_\chi) ]$ for the two-(three-)$\varphi$ annihilation mode. Heavier mediators typically bend the spectrum to be slightly harder.
Each plot is shown with a fixed benchmark annihilation cross-section. The spectrum scales linearly with this cross-section $\langle\sigma v\rangle$ and quadratically with the local dark matter density $\rho_\odot$. To aid in rescaling estimates, we also provide the cross-sections for a contracted NFW profile $\gamma_\text{\tiny NFW}=1.2$ which produce the same curves. 
Note that the contracted profile only contains a rescaling by the $J$-factor. Because the interstellar radiation field is not uniform in the region of interest, it is possible that the contracted profile  may lead to a change in the ICS spectrum.
The range of cross-sections can be interpreted as an estimate of uncertainty when comparing to the thermal relic cross-section.

For comparison to the observed Fermi-LAT spectrum, we plot in grey the systematic error band of the $\gamma$-ray excess spectrum as defined in the Fermi collaboration study of the $\gamma$-ray emission from the galactic center~\cite{TheFermi-LAT:2015kwa}. 
The Fermi collaboration provides two different estimates of the excess $\gamma$-ray spectrum and its systematic uncertainty:
\begin{enumerate}
	\item The first fits the excess as a parameterized exponential cutoff spectrum across the entire energy range of the data. This is shown in Fig.~\ref{fig:GCEspectra_parameterized}.
	\item The second fits the $\gamma$-ray  spectrum in independent energy bins. This is shown in Fig.~\ref{fig:GCEspectra_sys_eng_bin}.
\end{enumerate}

\paragraph{Flavor-dependence of $\gamma$-ray spectra.}
The $\mu e$ final states result in much harder $\gamma$-ray spectra than $\tau e$ or $\tau\mu$ final states. In the case of $\mu e$ final states, the ICS contributes the majority of the $\gamma$-ray flux at lower, $\mathcal O(1~\text{GeV})$, energies while the prompt contribution dominates at higher, $\mathcal O(10~\text{GeV})$, energies. 
In contrast, the ICS $\gamma$-ray flux in  the cases of $\tau e$ and $\tau\mu$ final states only constitutes a small fraction of the low energy spectrum, while the total signal is dominated by the prompt photon `bump',  which peaks between $2-10$~GeV before a spectral cutoff. This is because the hadronic $\tau$-decays allow for these annihilation channels to produce a much higher flux of prompt photons. 

\paragraph{Comparison to Fermi spectra.}
Our models are able to reasonably reproduce the parameterized Fermi $\gamma$-ray excess spectrum in Fig.~\ref{fig:GCEspectra_parameterized}.
The $\tau e$ and $\tau\mu$ model achieve this with dark matter masses of $m_\chi \sim 20-40~\text{GeV}$, while for the $\mu e$ case, slightly higher masses of $m_\chi\sim 40-100~\text{GeV}$ are required. 
The $\gamma$-ray excess is primarily produced through prompt emission from $\tau$ decays in the $\tau \mu$ and $\tau e$ models, whereas it is primarily produced through ICS in the $\mu e$ model. The dark matter mass for $\mu e$ final states must therefore be higher than the $\tau e / \tau \mu$ cases in order for the resulting electron spectrum to be hard enough to produce the Fermi $\gamma$-ray spectrum through ICS.
We note that although the $\gamma$-ray spectra produced in the $\mu e$ models may lie within the systematic error band defined by the Fermi collaboration's parametric fits, they are generally harder at high energies and do not have the characteristic peak at $\sim2-4$~GeV that is typically found in template analyses of the $\gamma$-ray excess.
The `cinched' shapes of the enveloped range of $\gamma$-ray emission in the $\tau e$ channels of Figs.~\ref{fig:GCEspectra_parameterized}--\ref{fig:GCEspectra_sys_eng_bin} arise due to a sampling effect: the spectra of the outgoing $\tau$'s and $e$'s do not change significantly until the mediator mass approaches its allowed minimum (the heavier lepton mass). As the enveloped region is defined using four linearly spaced values of $m_\phi$, the spectra for the three heavier mediator masses in the $\tau e$ case are very similar to each other. 

On the other hand, Fig.~\ref{fig:GCEspectra_sys_eng_bin} shows that our computed spectra are worse at fitting the Fermi collaboration's $\gamma$-ray spectra obtained through bin-by-bin fits. 
The spectrum derived through fitting the data in individual energy bins displays an extended, power-law-like tail at energies $\ge 10~\text{GeV}$; this was observed in \cite{Calore:2014nla, Calore:2015nua} and has recently been explored further in \cite{Horiuchi:2016zwu}. Our theoretical $\gamma$-ray spectra all cut off sharply around $\mathcal O(m_\chi)$ and thus cannot reproduce this spectral feature. 
One can interpret the difference between the parameterized and bin-by-bin fits as a qualitative assessment of the uncertainty in the target region for the spectral fit.

\paragraph{Compatibility with relic abundance.} 
For $\tau e$ and $\tau\mu$ final states, we are able to produce the observed Galactic Center excess flux with an annihilation cross-section $\langle \sigma v \rangle$ roughly equal to the relic density cross-section, (\ref{eq:relic:abundance:xsec}). 
For $\mu e$ final states, the annihilation cross-section must be $3-4$ times higher than the canonical relic cross-section in order to match the intensity of the Fermi Galactic Center excess. 
This is again related to the fact that there are substantially fewer prompt photons in the $\mu e$ scenario, hence the larger annihilation rate needed to account for the excess. 
We point out, however, the $\mu e$ states are brought back into consistency with (\ref{eq:relic:abundance:xsec}) if one instead invokes a contracted NFW profile. 

\paragraph{Dwarf spheroidal bounds.} Dwarf spheroidals are satellite galaxies that are rich in dark matter but have relatively little stellar matter. As a result, they typically set the strongest bounds on models of the galactic center excess that rely on prompt photon emission~\cite{Ackermann:2015zua}. This is avoided when the $\sim$GeV photons are produced though the inverse scattering of starlight because the dwarfs have a weak interstellar radiation field. Thus the $\mu e$ models are able to completely evade the dwarf bounds. 

For $\tau e$ and $\tau \mu$ channels, the decay into two differently flavored leptons means that the cross-section for annihilation to $\tau$'s is half of the total annihilation cross-section. We also note that the range of dark matter particle masses considered here for annihilations to mediators are generally higher than the best-fit masses in the case of direct annihilation. This is because either four or six SM leptons are produced per annihilation in these models instead of two in the direct scenario. The dwarf constraints on the annihilation cross-section into $\tau$'s are roughly $\sim1.5-2$ times weaker at 
$m_\chi \sim 20-40$ GeV  compared to $m_\chi \sim 10~\text{GeV}$, which is often quoted as the best-fit mass for direct annihilations into $\tau$'s. These two effects combine to partially alleviate the existing tensions with dwarf constraints on prompt $\gamma$-ray flux from  annihilation into $\tau$'s; the cross-sections required in our model for $\tau e$ and $\tau \mu$ channels are within $1\sigma$ of the limits from Ref.~\cite{Ackermann:2015zua}.

\paragraph{Comparison to Kaplinghat et al.}
We briefly compare our results to the $\chi\bar\chi \to VV \to 4e$ scenario; in doing so, we may highlight the differences in the lepton-flavor violating case and the role of uncertainties in astrophysical parameters.  
Kaplinghat et al.~\cite{Kaplinghat:2015gha} found that the annihilation mode to $4e$ fits the Galactic Center excess for an annihilation cross-section of $\langle \sigma v \rangle \approx \langle \sigma v \rangle_\text{rel.}/7$. By comparison, our $2(\mu e)$ annihilation mode is found to fit with $\langle \sigma v \rangle \approx 3 \langle \sigma v \rangle_\text{rel.}$. The difference between these factors are predominantly from three sources:
\begin{enumerate}
	\item Our result uses a standard $\gamma_\text{\tiny NFW}=1.0$ dark matter halo profile while Kaplinghat et al.\ use a contracted $\gamma_\text{\tiny NFW}=1.2$ profile. As shown in the plot, the difference in $J$ factors is approximately a factor of 3.
	\item The discussion in Sec.~\ref{sec:photons:from:leptons} shows that electrons and muons produce very different ICS spectra. This is in contrast to prompt photon production where the mass difference can be ignored at sufficiently high energies. This introduces an $\mathcal O(\text{few})$ difference between the flavor violating and flavor conserving modes.
	\item Finally, there are some $\mathcal O(\text{few})$ differences in the target region, Kaplinghat et al.\ use the fit by \cite{Calore:2014xka} while we use the envelopes from the Fermi collaboration~\cite{TheFermi-LAT:2015kwa}.
\end{enumerate}

\section{The AMS-02 $e^\pm$ Spectrum}
\label{sec:ams}

%%%%%%%%%%%%%%%%%%%%%%%%%%%%%%%%%%%%%%%%%%%%%%%%%%%%%%%%%%
\begin{figure}%[h!]
\centering
\begin{subfigure}{0.45\textwidth}
\includegraphics[width=\textwidth]{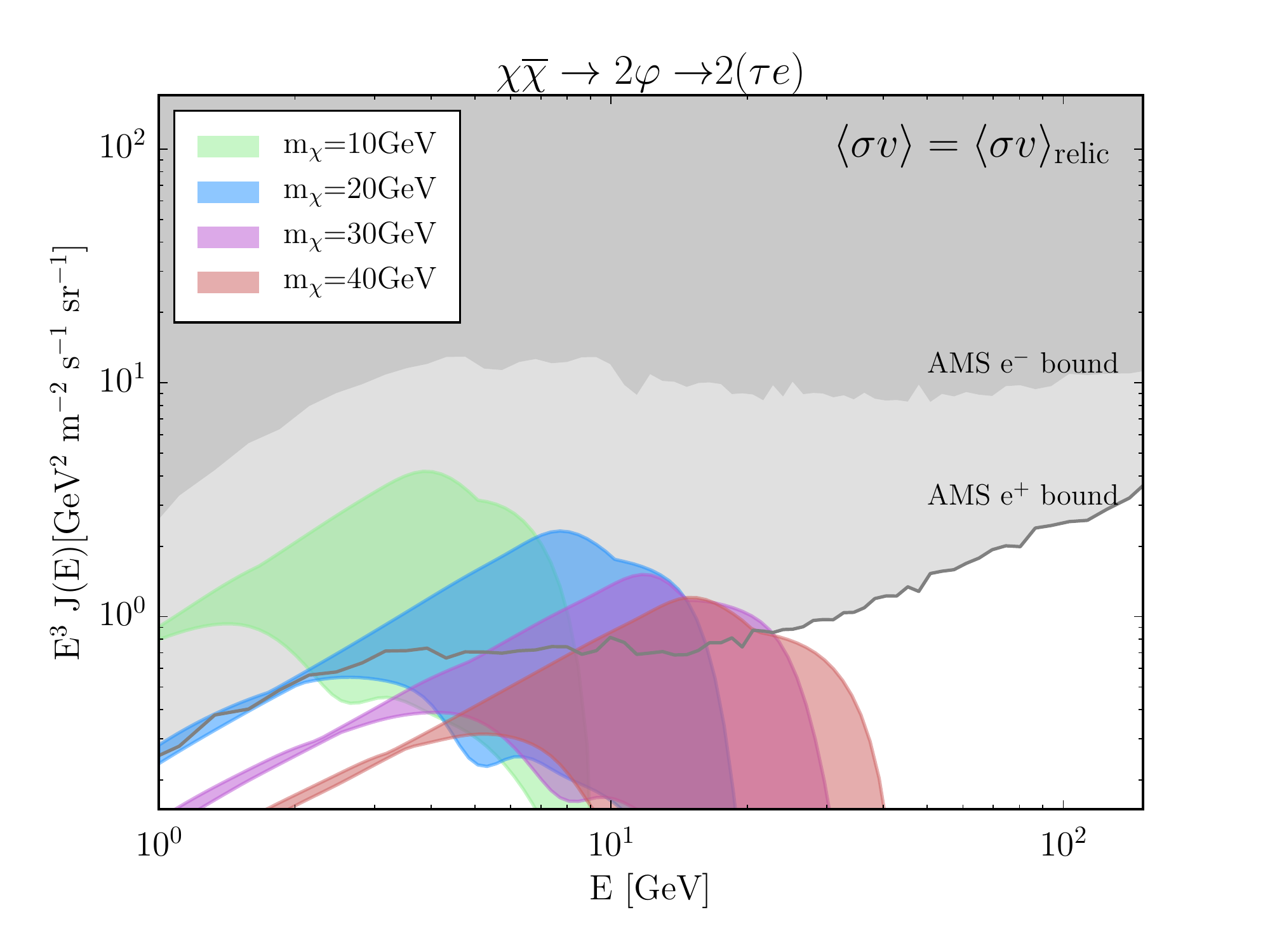}
%\caption{$ $}
\label{fig:}
\end{subfigure}
\begin{subfigure}{0.45\textwidth}
\includegraphics[width=\textwidth]{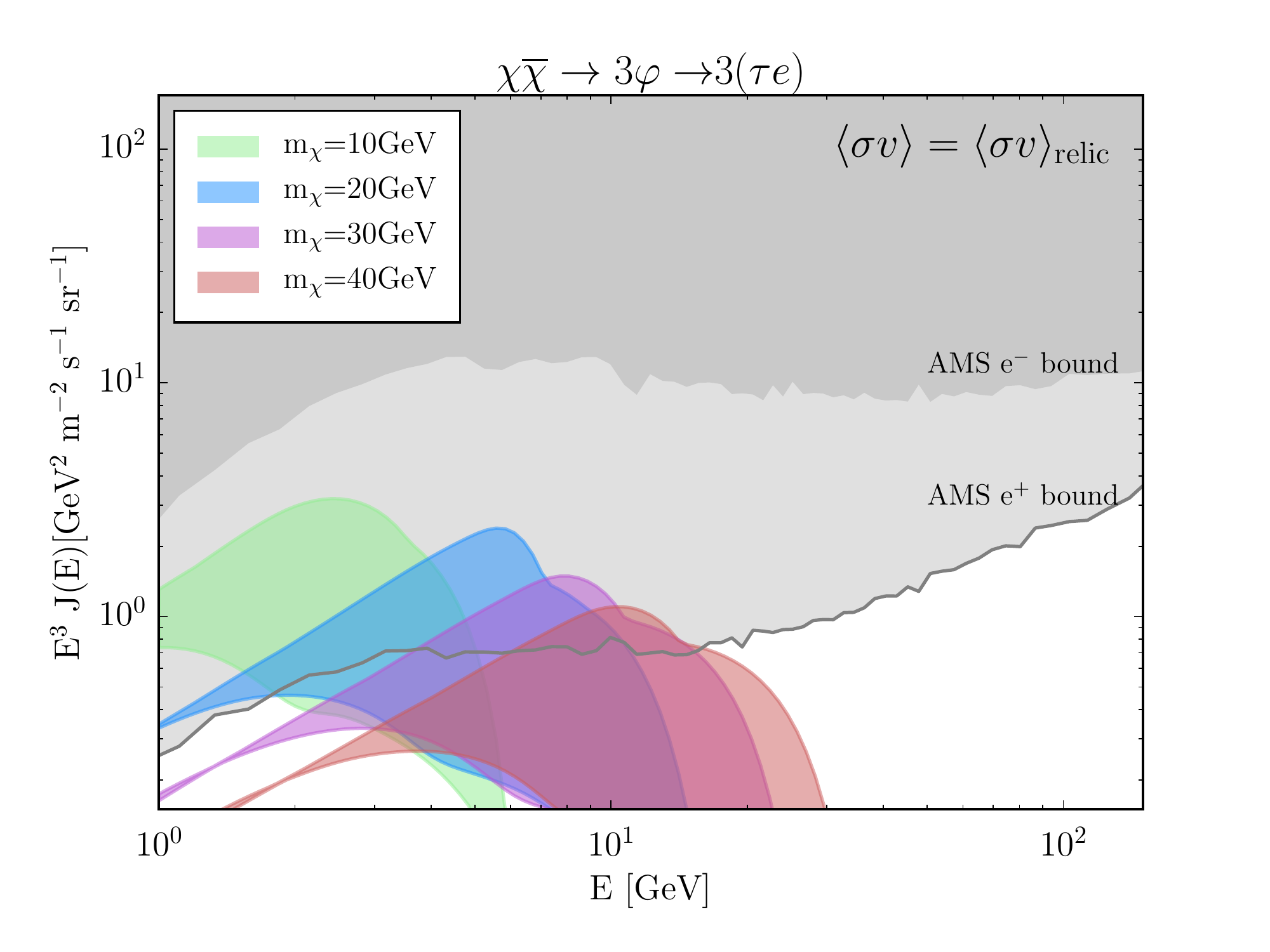}
%\caption{$ $}
\label{fig:}
\end{subfigure}
\begin{subfigure}{0.45\textwidth}
\includegraphics[width=\textwidth]{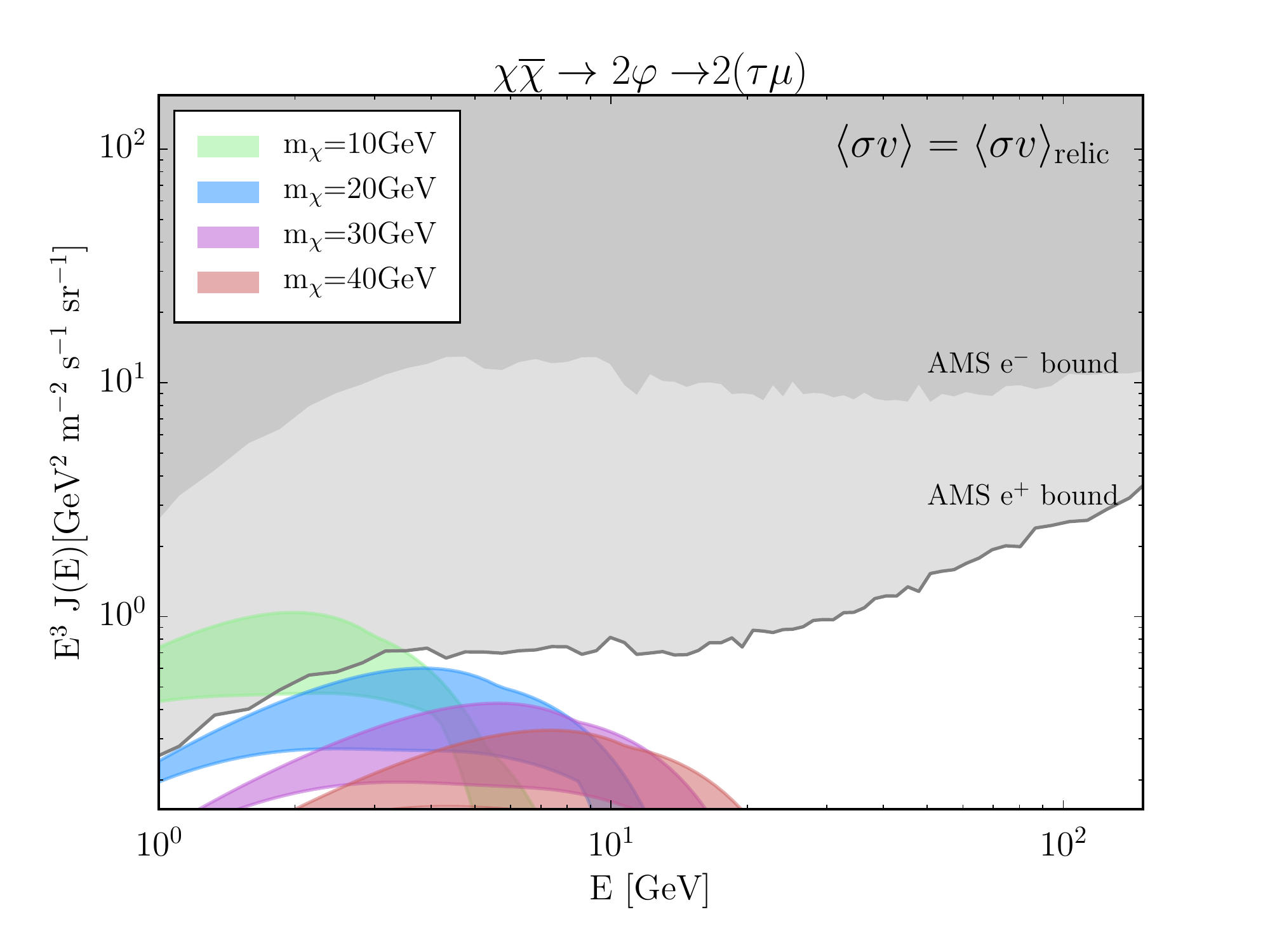}
%\caption{$ $}
\label{fig:}
\end{subfigure}
\begin{subfigure}{0.45\textwidth}
\includegraphics[width=\textwidth]{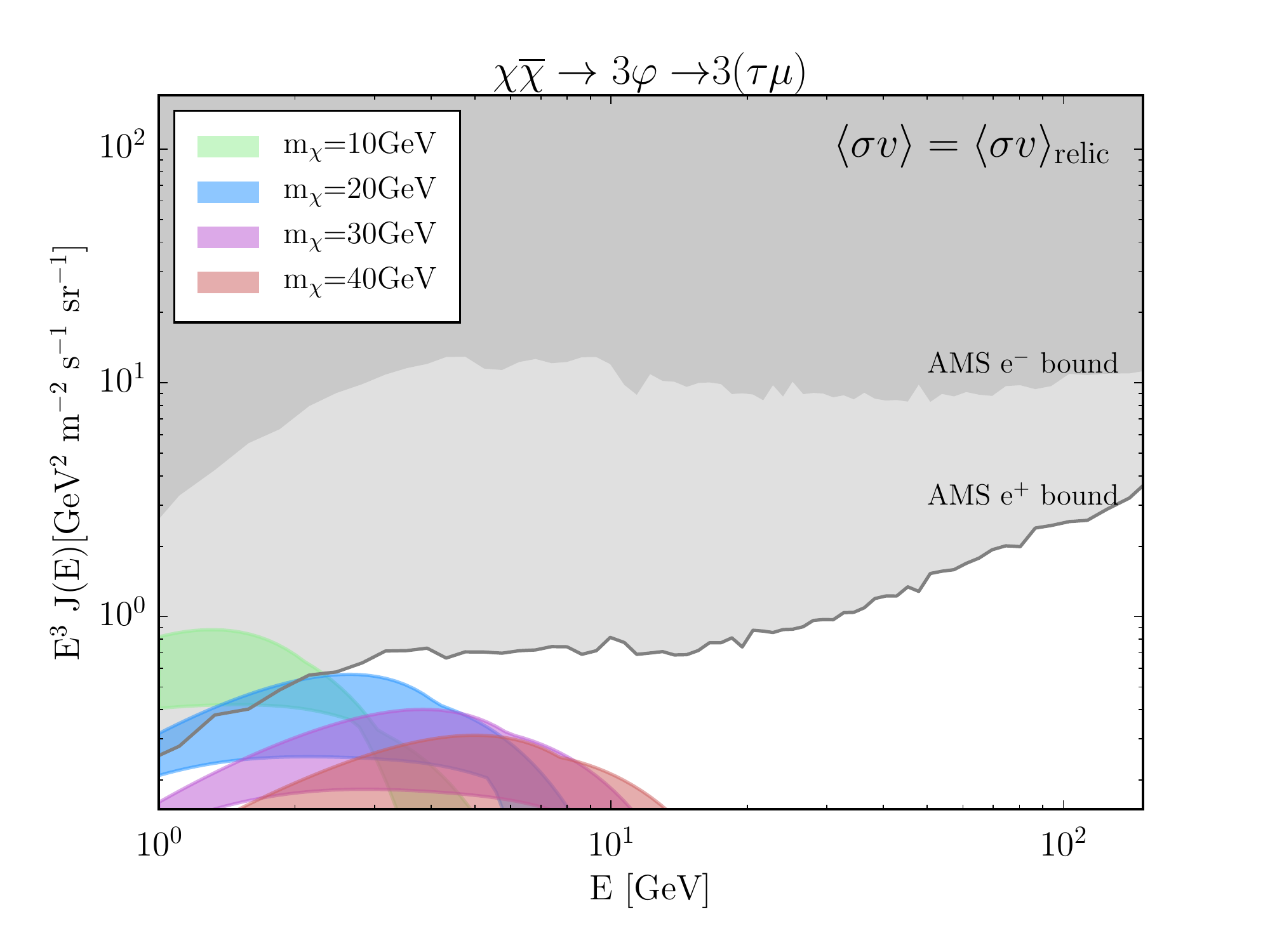}
%\caption{$ $}
\label{fig:}
\end{subfigure}
\begin{subfigure}{0.45\textwidth}
\includegraphics[width=\textwidth]{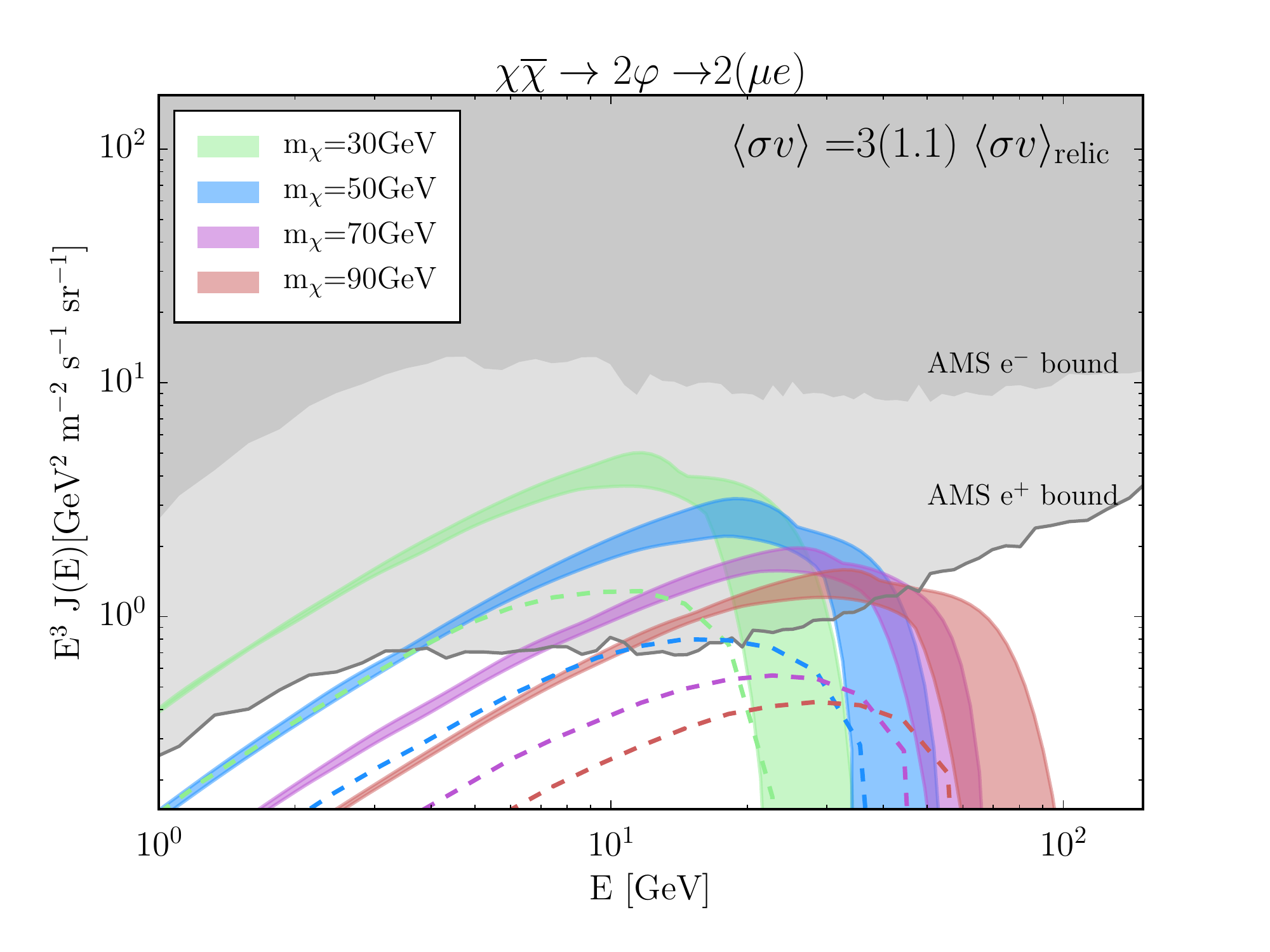}
%\caption{$ $}
\label{fig:}
\end{subfigure}
\begin{subfigure}{0.45\textwidth}
\includegraphics[width=\textwidth]{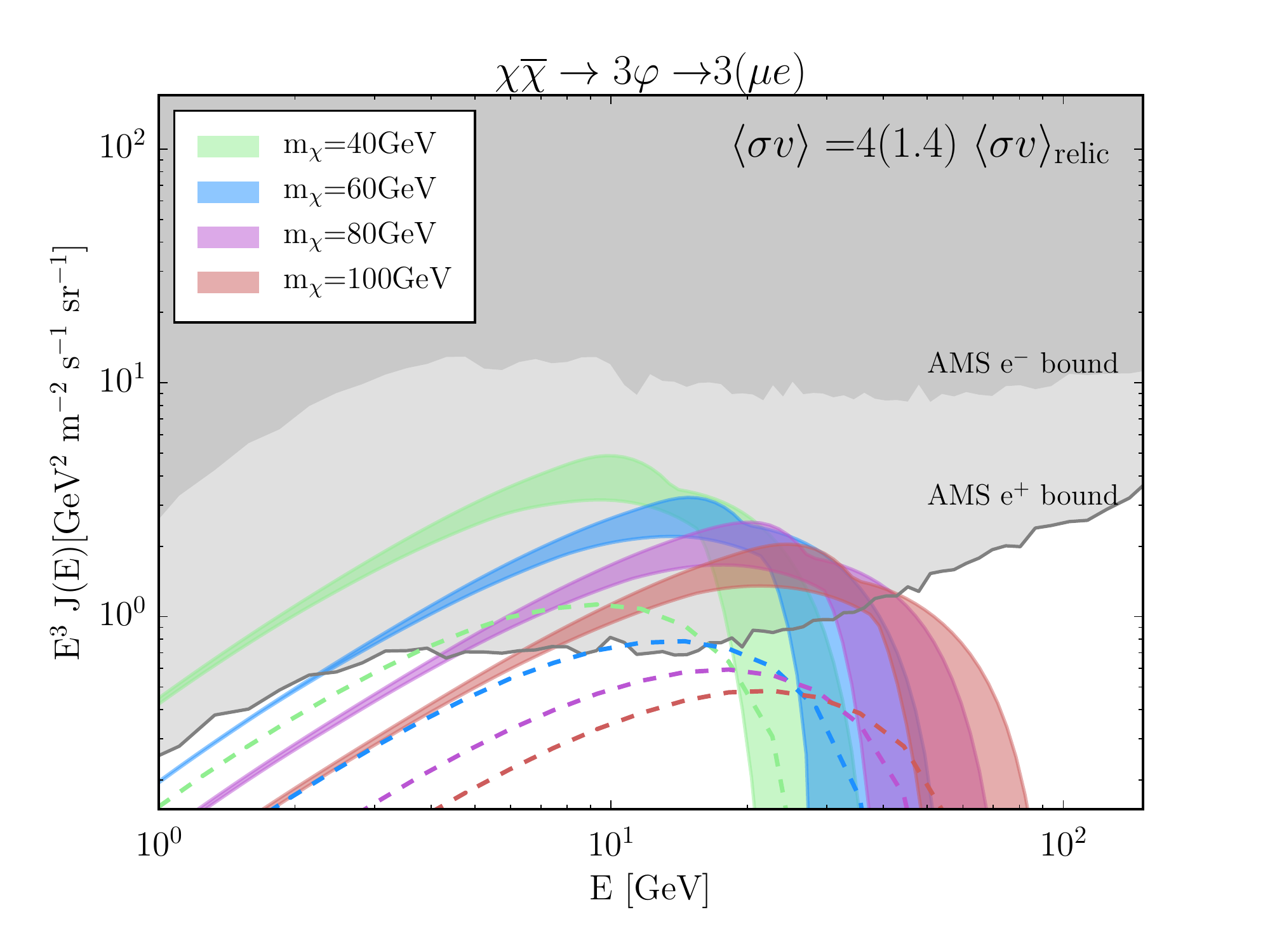}
%\caption{$ $}
\label{fig:}
\end{subfigure}
\caption{
Predicted $e^{\pm}$ spectra at Earth's position for each scenario: $\chi\bar\chi\to2\varphi$ (left) and $\chi\bar\chi\to3\varphi$ (right) followed by one the lepton-flavor violating decays indicated in each panel. The grey shaded region represents the bounds from AMS-02 electron and positron spectra while each color-coded band corresponds to a set of $\{m_\chi,~m_\phi \}$ with $m_\phi$ varying in the range
$[m_{\ell_{\text{heavy}}},~m_\chi]$ (left) and $[m_{\ell_{\text{heavy}}},~\frac23 m_\chi]$ (right).
The dashed lines shown in the bottom row correspond to the prediction for AMS-02 spectra if an NFW profile slope of $\gamma_\text{\tiny NFW}=1.2$ is assumed. This leads to a lower annihilation cross section of 1.1 (1.4) times the relic density for $\chi\bar \chi \to 2(3) \varphi$.
}
\label{fig:AMSbounds}
\end{figure}
%
%%%%%%%%%%%%%%%%%%%%%%%%%%%%%%%%%%%%%%%%%%%%%%%%%%%% 

The AMS-02 experiment may be capable of detecting electrons and positrons produced in dark matter annihilations~\cite{Bergstrom:2013jra}. AMS observations thus far have found that the $e^{\pm}$ spectra are smoothly varying, with no line-like or bump features~\cite{Aguilar:2014mma}. We constrain the range of particle properties allowed in our model by requiring that the flux of $e^{\pm}$ produced through annihilations and propagated to Earth's position in the Milky Way must be low enough to avoid producing any such features in the observed $e^{\pm}$ spectra. To do so, we use the \texttt{DRAGON} 3D cosmic ray propagation code along with the propagation setup described in Ref.~\cite{Gaggero:2013rya}. The diffusion coefficient is assumed to depend on particle rigidity as 
\begin{align}
 D(\rho)&=\beta^{-0.4}D_{0} \left(\frac{\rho}{\rho_{0}}\right)^{\delta}
 &
\text{with \;} 
D_{0}=3\times10^{28}~\text{cm}^{2} \text{s}^{-1}\, , 
\;\rho_{0}=3 \text{GV}\, ,
\;\delta=0.6 \, .
\end{align}
We assume the same halo parameters and annihilation cross-section for each case of lepton final states and number of mediators as in the previous section.

Fig.~\ref{fig:AMSbounds} shows the expected $e^\pm$ energy spectrum from $\chi\bar\chi$ annihilation to $\varphi$ followed by the decay $\varphi\to\bar\ell_i\ell_j$.
Following the notation of the previous section, results are presented for each annihilation mode and each of the three $\varphi$ decay models separately.
We fix the dark matter annihilation rate of each model to roughly match the observed Galactic Center $\gamma$-ray excess flux.
We regard a model as consistent with the AMS observations if for all energies, the $e^\pm$ flux predicted by the model is lower than the total size of the error bars given in Ref.~\cite{Aguilar:2014mma} for the $e^{+}$ and $e^{-}$ binned fluxes at that energy. 

We find that for $\tau e$ and $\tau\mu$ final states, dark matter masses above $\sim 20\text{ GeV}$ are not excluded by AMS observations, and can also produce a $\gamma$-ray signal consistent with the Fermi excess. 
For $\mu e$ final states, all potential e$^{\pm}$ spectra studied here are in tension with the AMS positron bounds when $\gamma_\text{\tiny NFW}=1.0$. This tension may be reduced by either
\begin{enumerate}
	\item increasing the dark matter mass 
beyond 90--100 GeV, or
	\item considering of a steeper dark matter halo density profile.
\end{enumerate}
Increasing the mass comes at the expense of hardening the $\gamma$-ray signal and introduces tension with the Fermi result. On the other hand, a slightly steeper NFW inner profile slope or $\gamma_\text{NFW}=1.1-1.3$ was suggested in \cite{Daylan:2014rsa, Calore:2014xka} for the Fermi $\gamma$-ray excess. When using a steeper profile, the annihilation cross-section needed to reproduce the observed excess brightness decreases by up to a factor of $\sim 5$.
Since the local dark matter density is held fixed as the density profile slope changes, this lower annihilation rate results in a decrease in the dark matter contribution to the {AMS} $e^\pm$ spectrum. The potential dark matter contribution to the local $e^\pm$ spectrum is dominated by the flux from annihilations near the solar neighborhood; changes to the profile near the galactic center have little effect on this measurement.
The dashed lines in the $\mu e$ plots of Fig.~\ref{fig:AMSbounds} show the upper contour of the $e^\pm$ spectra for the estimated reduced annihilation cross-sections with a contracted $\gamma_\text{\tiny NFW}$ profile. One can see that the factor of $\sim 3$ in the cross-section allows these modes to avoid the AMS $e^+$ bound.

\section{Constraints on Standard Model Couplings}
\label{eq:constraints}

The discussions in Sections~\ref{sec:fermi}--\ref{sec:ams} focused on the target region and constraints on the dark sector couplings, (\ref{eq:L:phi:chi}), with the assumption that the mediator decays are sufficiently prompt on astrophysical scales. In this section we review the constraints on the Standard Model couplings, (\ref{eq:L:phi:SM}), that control that decay length. We emphasize that by virtue of the hidden sector scenario, these couplings can be taken to be very small to avoid the bounds here without causing the decay length to distort the Fermi or AMS analyses above.

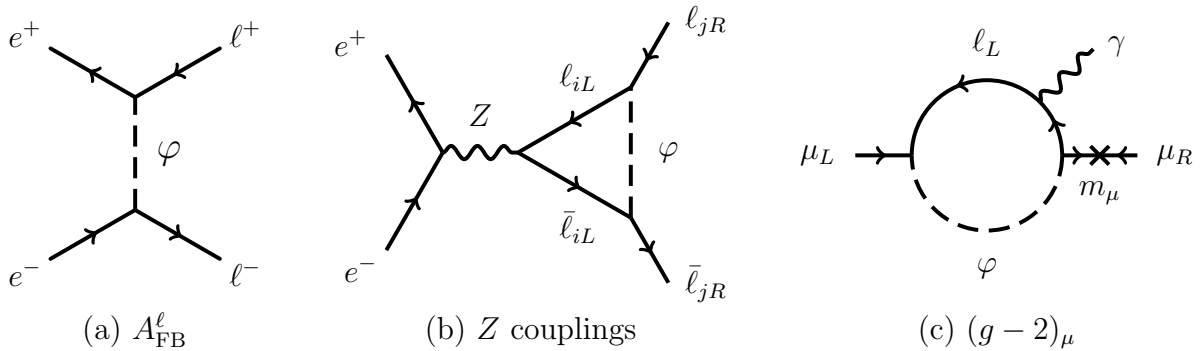
\begin{figure}
\centering
\begin{tabular}{ccc}
	\raisebox{-.5\height}{
		\begin{tikzpicture}[line width=1.5 pt]
% Just rotate the previous one and make a few tweaks
	\def\labelscaling{0.3}		% distance of label from endpoint
	\def\labelscalingup{-0.45}	% distance of label from line
	\def\legangle{60}			% angle of legs
	\def\leglength{1.3}			% length of legs
\begin{scope}[rotate=90]
	\coordinate (v1) at (-.75,0); 	% left vertex
	\coordinate (v2) at (.75,0); 	% right vertex
	\coordinate (vhalf) at ($ (v1) !.5! (v2) $); % middle
	\coordinate (leg1a) at ($(v1)-( \legangle:\leglength)$); % bot left endpoint
	\coordinate (leg1b) at ($(v1)-(-\legangle:\leglength)$); % top left endpoint
	\coordinate (leg2a) at ($(v2)+( \legangle:\leglength)$); % top right endpoint
	\coordinate (leg2b) at ($(v2)+(-\legangle:\leglength)$); % bot right endpoint
%	
	% label positions
	\coordinate (l1a) at ($(leg1a) - \labelscaling*( \legangle:\leglength)$); 
	\coordinate (l1b) at ($(leg1b) - \labelscaling*(-\legangle:\leglength)$);
	\coordinate (l2a) at ($(leg2a) + \labelscaling*( \legangle:\leglength)$);
	\coordinate (l2b) at ($(leg2b) + \labelscaling*(-\legangle:\leglength)$);
	\coordinate (lc) at ($(vhalf)+\labelscalingup*(0,1)$);
%
	% draw topologies
	\draw[dashed, dash pattern=on 8 off 4] (v1)--(v2);
	\draw[fermion, line cap=round] (v1)--(leg1a);
	\draw[fermion, line cap=round] (leg1b)--(v1);
	\draw[fermion, line cap=round] (v2)--(leg2a);
	\draw[fermion, line cap=round] (leg2b)--(v2);
\end{scope}
%	
	% draw nodes
	\node at (l1a) { $\displaystyle \ell^-$};
	\node at (l1b) { $\displaystyle e^-$};
	\node at (l2a) { $\displaystyle e^+$};
	\node at (l2b) { $\displaystyle \ell^+$};
	\node at (lc) {\large $\varphi$};
\end{tikzpicture}
	}
	&
	\raisebox{-.5\height}{
		\begin{tikzpicture}[line width=1.5 pt]
	\begin{scope}[shift={(-.5,0)}]
		\draw[fermionbar] (120:1.5) -- (0,0);
		\draw[fermionbar] (0,0) -- (240:1.5);
		\node at (130:1.9) { $\displaystyle e^+$};
		\node at (235:1.9) { $\displaystyle e^-$};
	\end{scope}
		\draw[vector] (-.5,0) -- (0.5,0);
		\node at (0,.5) { $\displaystyle Z$};
	\begin{scope}[shift={(1.5,0)}] % Mass Insertion
%	\draw[vector] (180:1) -- (180:2);
	\draw[fermionbar] (60:1) -- (60:2);
	\draw[fermion] (300:1) -- (300:2);
	\node at (-.2,1) { $\displaystyle \ell_{iL}$};
	\node at (-.2,-1) { $\displaystyle \bar\ell_{iL}$};
	\draw[fermion] (180:1) -- (300:1);
	\draw[dashed, dash pattern=on 8 off 4]  (300:1) -- (60:1);
	\draw[fermion] (60:1) -- (180:1);
	\node at (1,0) { $\displaystyle \varphi$};
	\node at (1.5,-1.75) { $\displaystyle \bar\ell_{jR}$};
	\node at (1.5,+1.75) { $\displaystyle \ell_{jR}$};
	\end{scope}
 \end{tikzpicture}
	}
	&
	\raisebox{-.5\height}{
		\begin{tikzpicture}[line width=1.5 pt, scale=1]
		\draw[fermion] (-1.75,0) -- (-1,0);
		\draw[fermion] (1,0) -- (1.5,0);
		\draw[fermion] (2,0) -- (1.5,0);
		\draw[fermionbar] (-1,0) arc (180:45:1);
		\draw[fermionbar] (45:1) arc (45:0:1); 
		\draw[dashed, dash pattern=on 8 off 4] (1,0) arc (0:-180:1);
		\draw[vector] (45:1) -- (45: 2);
	\node at (0,-1.5) {$\varphi$};
	\node at (-2.25,0) {$\mu_L$};
	\node at (2.5,0) {$\mu_R$};
	\node at (1.5,-.5) {$m_\mu$};
	\node at (0,1.5) {$\ell_L$};
	\node at (40:2.25) {$\gamma$};
	\begin{scope}[shift={(1.5,0)}] % Mass Insertion
		\clip (0,0) circle (.175cm);
		\draw[fermionnoarrow] (-1,1) -- (1,-1);
		\draw[fermionnoarrow] (1,1) -- (-1,-1);
	\end{scope}	
	\end{tikzpicture}
	}
	\\
	(a) $A_\text{FB}^\ell$ 
	& (b) $Z$ couplings
	& (c) $(g-2)_\mu$
\end{tabular}
	\caption{Diagrams demonstrating possible constraints on the Standard Model couplings of the $\varphi$, (\ref{eq:L:phi:SM}). In (c), arrows represent helicity to show that an external mass insertion is required.
	}
	\label{fig:SM:coupling}
\end{figure}

\subsection{Photon lines}

Mediators may decay into photon pairs, $\varphi \to \gamma \gamma$, if flavor-conserving couplings are generated. These decays would be seen in the galactic $\gamma$-ray spectrum~\cite{Abazajian:2011tk}. Such couplings are assumed to be negligible as they are only generated by $(L_i-L_j)$-breaking effects.

\subsection{Electroweak Precision Measurements}

Electron colliders are able to probe the chiral structure of new physics through the forward--backward asymmetries of $e^+e^-\to f\bar f$ scattering~\cite{langacker1996precision},
\begin{align}
	A_\text{FB}^f &= \frac{\sigma_>(e^+e^-\to f\bar f) - \sigma_<(e^+e^-\to f\bar f)}{\sigma_>(e^+e^-\to f\bar f)+\sigma_<(e^+e^-\to f\bar f)}\ ,
\end{align}
where $\sigma_{>(<)}$ refers to the forward (backward) cross-section where the azimuthal angle of the $f$ with respect to the $e^-$ has positive (negative) cosine. 
The $t$-channel exchange of a mediator modifies the forward--backward asymmetry relative to its Standard Model value; this is shown in Fig.~\ref{fig:SM:coupling}a. Observations of $A_\text{FB}^\ell$ for $\ell = \mu,\tau$ therefore constrain the couplings and mass of $\varphi$. 
It is straightforward to translate such bounds on supersymmetric $R$-parity violating models to our scenario, by identifying the sneutrino with
$\varphi$, and decoupling the rest of the supersymmetric spectrum.
Comparing with~\cite{Barger:1989rk, Barbier:2004ez}, we find
\begin{align}
	g_{\mu e, e \mu} & < 2.5\times 10^{-3} \left(\frac{m_\varphi}{\text{GeV}}\right)
	&
	g_{\tau e, e \tau} & < 1.1\times 10^{-3} \left(\frac{m_\varphi}{\text{GeV}}\right)
	\ .
	\label{eq:bound:FB}
\end{align}

The chiral couplings of the $Z$ boson are also precisely measured by SLD and LEP through $e^+e^- \to Z\to \ell^+\ell^-$~\cite{ALEPH:2005ab}. In our model, vertex corrections with an internal $\varphi$ line will mix the left-handed $\ell_j$ and right-handed $\ell_i$ couplings; this is shown in Fig.~\ref{fig:SM:coupling}b. Altmannshofer et al.\ recently performed full analysis of these couplings, including error correlations in the couplings, for the case a $\tau\mu$ lepton-flavor violating spin-1 boson~\cite{Altmannshofer:2016brv}. They found that the bound from this measurement is typically much weaker than the anomalous magnetic moment of the muon. We thus assume that these constraints are subdominant to (\ref{eq:bound:FB}) for the $\mu e$ and $\tau e$ couplings and the muon magnetic moment (discussed below) for the $\tau \mu$ couplings of $\varphi$.

Other subdominant constraints include corrections to the $Z$ and $W$ widths from the on-shell emission of $\varphi$ off a charged lepton decay product~\cite{Laha:2013xua}, corrections to the Peskin--Takeuchi parameters (which begin at two-loop order)~\cite{Peskin:1991sw}, and contribution to the highly suppressed multi-lepton decay modes of charged kaons.

\subsection{Lepton Anomalous Dipole Moments}

The anomalous electric- and magnetic-dipole moments of leptons, place strict constraints on light new physics~\cite{Pospelov:2005pr,Pospelov:2008zw, Davoudiasl:2012ig, Endo:2012hp,Giudice:2012ms}.
The interaction structure of (\ref{eq:L:phi:SM}), assures
that with a single chiral coupling, a complex phase
in $g_{ij}$ can be rotated away in the $(L_i-L_j)$-symmetric limit.
As a result, electric dipole moments do not play a role in constraining the allowed parameter-space of the coupling.
On the other hand, contributions to magnetic dipole operators are generated already at the one-loop level, and are experimentally constrained. These operators involve both left- and right-lepton chirality states, and therefore require mass insertions for (\ref{eq:L:phi:SM}) to contribute.
The contribution of (\ref{eq:L:phi:SM}) to the anomalous magnetic dipole moment, $a_{\ell_j} = \frac12\left(g-2\right)_{\ell_j}$, is
\begin{align}
	\Delta a_{\ell_j} = 
	\frac{m_{\ell_j}}{16\pi^2}
	\sum_{i=1}^3
	\int_0^1 dx\,
	(1-x)^2
	\frac{x m_{\ell_j} S_i + m_{\ell_i} P_i }
	{x m_\varphi^2+(1-x)m_{\ell_i}^2-x(1-x)m_{\ell_j}^2}
	\ ,
	\label{eq:delta:a}
\end{align}
where $\Delta a_\ell = a_\ell^{\rm{exp}} - a_\ell^{\rm{SM}}$ is the deviation from the Standard Model prediction, $S_i = |g_{ij}|^2 + |g_{ji}|^2$, and $P_i = g^*_{ij}g_{ji} +  g^*_{ji}g_{ij}$; see also~\cite{Kim:2001se} where $\varphi$ is identified with a sneutrino with an $R$-parity-violating interaction.

\begin{figure}
\center
\includegraphics[width=0.32\textwidth]{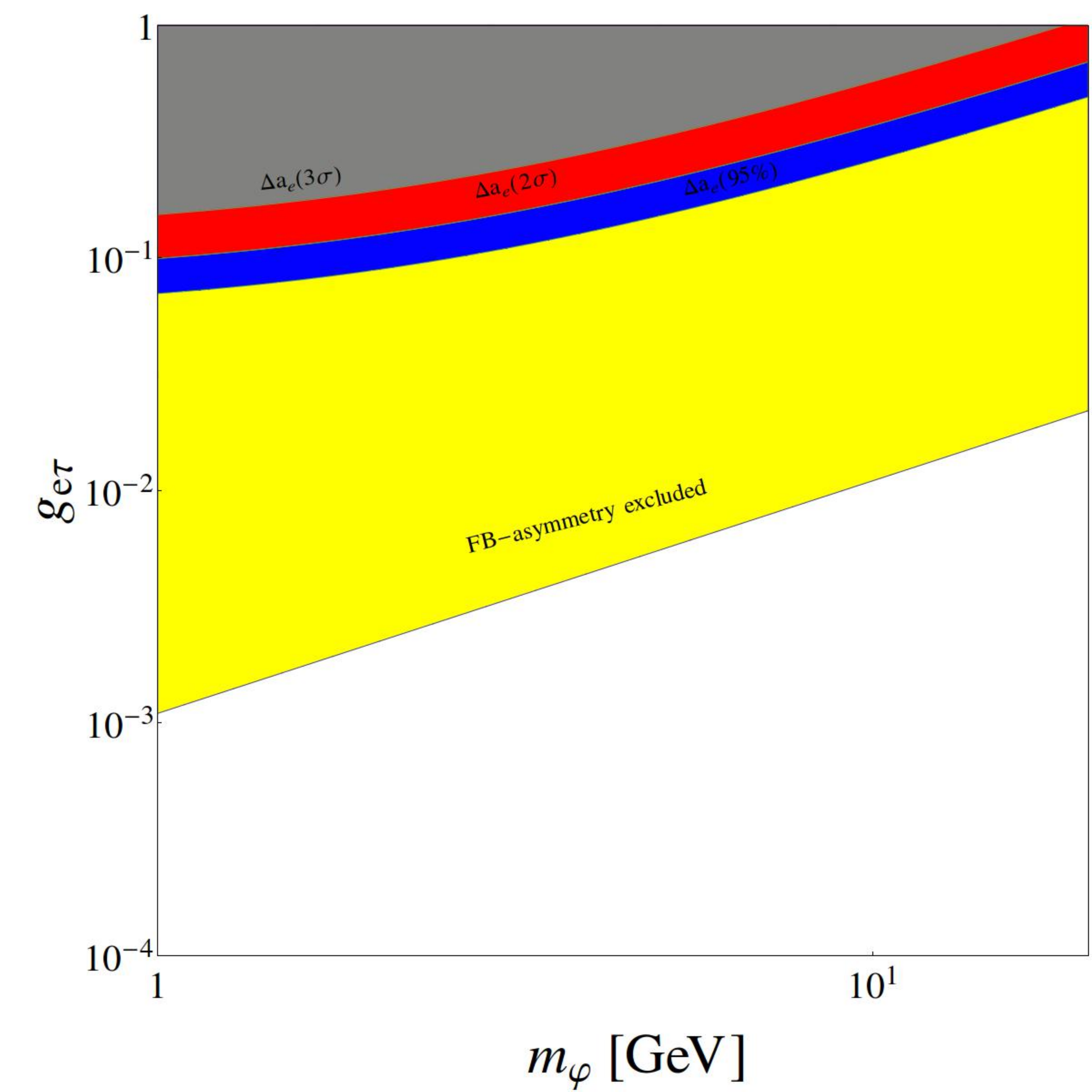}
\includegraphics[width=0.32\textwidth]{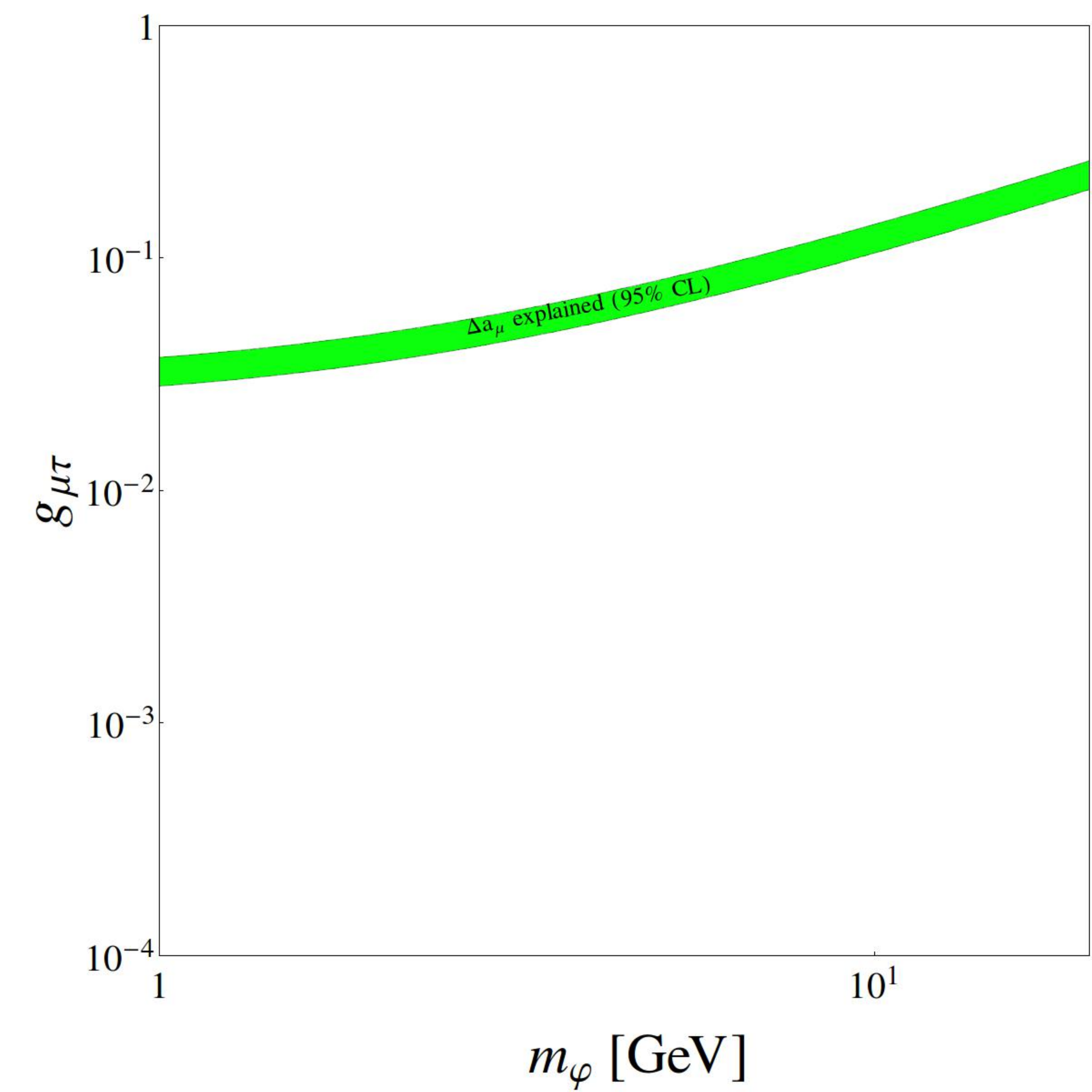}
\includegraphics[width=0.32\textwidth]{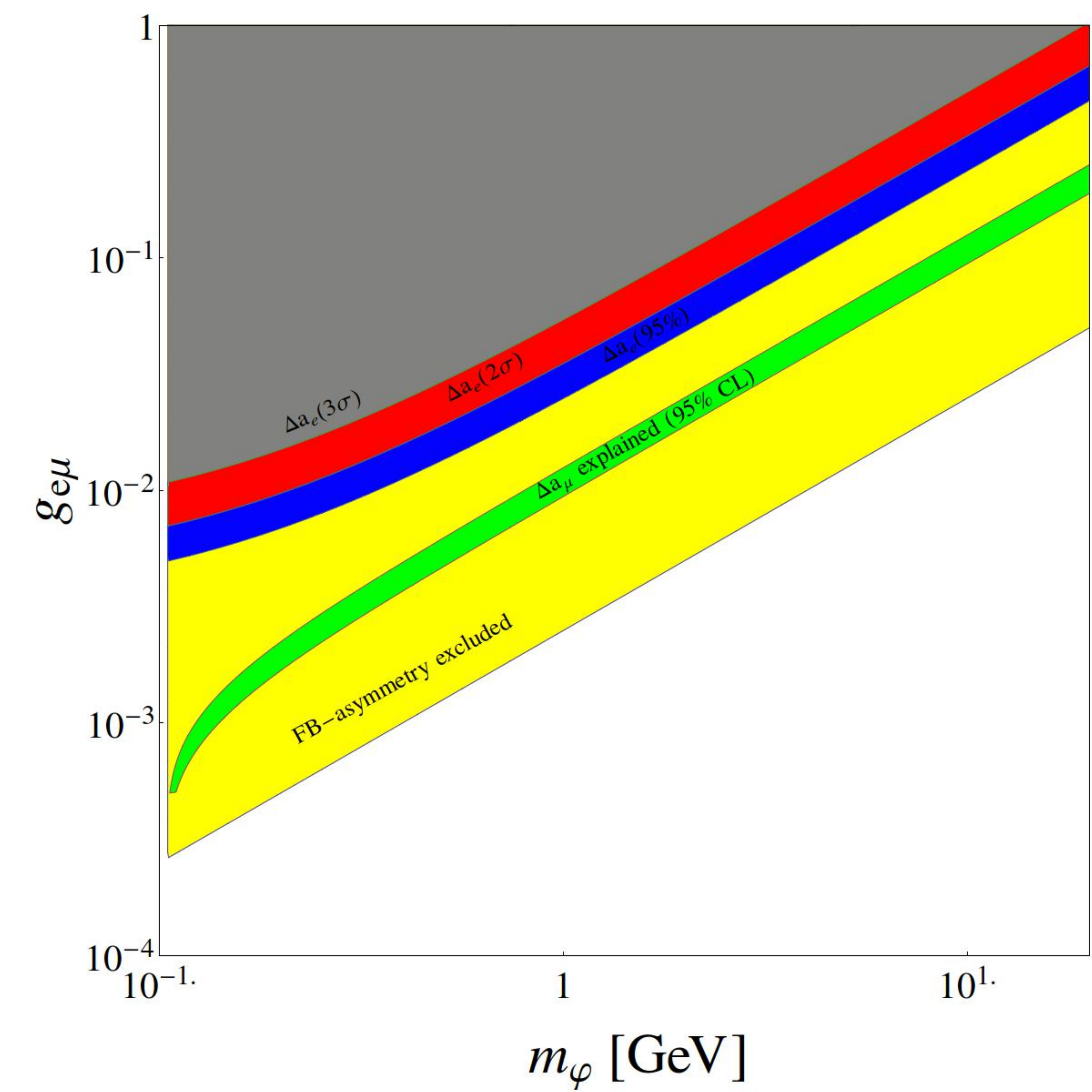}
\caption{Excluded and preferred parameter-space of the flavor-conserving observables: $\Delta a_e,~\Delta a_\mu$, and $A_\text{FB}^f$ in
the $g_{ij}-m_\varphi$ plane.
The entire $\Delta a_{e,\mu}$ excluded and preferred region 
falls into the domain excluded by the forward-backward asymmetry (yellow).
The $\mu\tau$ model has no constraint from the forward-backward asymmetry
and can account for both the $(g-2)_\mu$ excess and the Fermi $\gamma$-ray excess.
}
\label{fig:diag_flv_bounds}
\end{figure}

The bounds on the electron~\cite{Hanneke:2008tm, Aoyama:2012wj} and muon~\cite{Bennett:2006fi} magnetic moments are
\begin{align}
	\Delta a_e & = -1.05 (0.82) \times 10^{-12}
	&
	\Delta a_\mu &= 288 (80) \times 10^{-11}
	\ ,
	\label{eq:MDM_bounds}
\end{align}
Note that the central value of $\Delta a_e$ is negative, and cannot be
accounted for by (\ref{eq:delta:a}) under the single coupling $g_{ij}$ 
assumption.
The chirality flip required to generate the dipole operator can only occur on an external line, as in Fig.~\ref{fig:SM:coupling}c.
The bounds from 
magnetic dipole moments, 
and the forward-backward asymmetry
are plotted in Fig.~\ref{fig:diag_flv_bounds}.
Following~\cite{Davoudiasl:2014kua}, we plot the exclusion
bounds for $\Delta a_e$ for various confidence levels, in the
parameter-space region where it is positive. We also plot the 
preferred parameter-space region to account for 
the 2.6$\sigma$ anomaly in $(g-2)_\mu$~\cite{Bennett:2006fi}.

\subsection{Charged Lepton Flavor Violation}

The $L_i - L_j$ symmetry suppresses traditional signatures
of charged lepton flavor violation~\cite{Bernstein:2013hba}.
While the contributions to most tree-level processes are zero,
symmetry breaking effects can still enter through loop-induced
interactions of charged leptons with neutrinos and $W$-bosons.
These effects are, however, highly suppressed due
to the loop-nature, the $W$-mass, and the leptonic GIM-mechanism
involving the small neutrino masses.
\begin{itemize}
	\item \textbf{Lepton radiative decays, $\ell_i \to \ell_j \gamma$}~\cite{Adam:2013mnn,Aubert:2009ag}. The one-loop contribution to these flavor-changing dipole operators vanishes. The leading $\varphi$ contribution starts at three-loop order and is highly suppressed.
	\item \textbf{$\mu \to e$ conversions in heavy nucleus}.~\cite{Bertl:2006up,Dohmen:1993mp,comet,Abrams:2012er} Since $\varphi$ does not directly couple to nuclei, the leading contributions appear from the flavor-changing dipole operator so that $\text{Br}\left(\mu~\rm{Nuc} \to e ~\rm{Nuc}\right)\approx \alpha_{EM}\times \text{Br}\left(\mu\to e\gamma\right)$. This dipole, however, vanishes due to the chiral nature of the $\varphi$ interactions.
	\item \textbf{Multibody rare lepton decays, $\ell_i \to \ell_j \ell_k\bar\ell_k$}~\cite{Bellgardt:1987du,Hayasaka:2010np} \textbf{ and $\ell_i \to \ell_j \ell_k\bar\ell_k\nu\bar\nu$} ~\cite{Bertl:1985mw}.
	As $\varphi$ has no tree-level flavor-conserving interactions the first process cannot proceed via tree-level interactions. The coupling constraints of the preceding section then imply that any induced flavor-conserving coupling is negligible.
	  If one allows for neutrino final states as in the second process, tree-level $W$-interactions then violate $L_i-L_j$ and $\varphi$ can participate in mediating the second type processes. In this case, both $\varphi$ and an intermediate lepton state must go off-shell making these types of contributions highly suppressed. We have numerically verified that such contributions are negligible, even for $g_{ij}=1$, and for $m_\varphi \gtrsim m_{\ell_j}~(j>i)$.
	  	\item \textbf{Muonium oscillation, $M \leftrightarrow \bar M$, $M=\bar\mu e$}~\cite{Willmann:1998gd}. In the absence of flavor-conserving $\varphi$ couplings, muonium oscillations place strong bounds on $\mu e$ flavor violation. In our case, the $L_i-L_j$ symmetry forbids tree-level contributions and $\varphi$ interactions can only decorate the already highly suppressed one-loop SM contributions\footnote{
On the other hand, a real scalar cannot carry spurious $L_\mu - L_e$ charge, in which case the bound is rather strong~\cite{Kim:1997rr}
$\left.g_{\mu e, e \mu}\right|_{\varphi \in \mathbbm{R}}
< 4.4\times 10^{-4} \left(\frac{m_\varphi}{\text{GeV}}\right)$.
}.
\end{itemize}

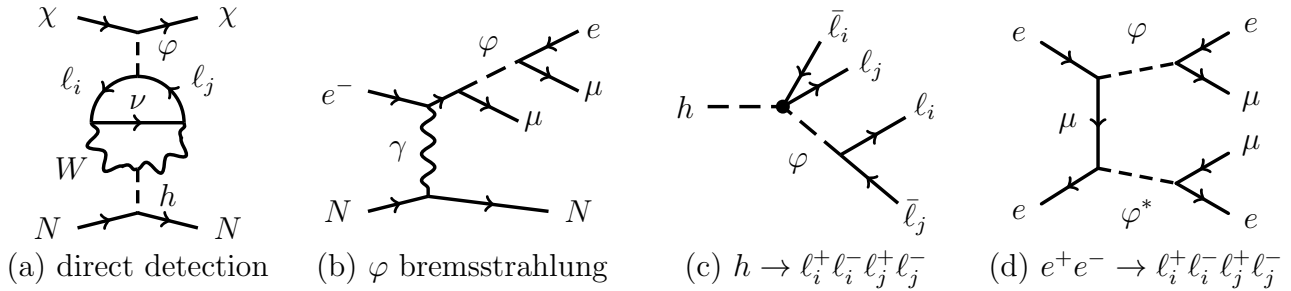
\begin{figure}
\centering
\begin{tabular}{cccc}
	\raisebox{-.5\height}{
		\begin{tikzpicture}[line width=1.3 pt, scale=.8]
	\def\radu{.77}
	\draw[fermion] (-1,1.75) -- (0,1.5);
	\draw[fermion] (0,1.5) -- (1,1.75);
	\draw[fermion] (-1,-1.75) -- (0,-1.5);
	\draw[fermion] (0,-1.5) -- (1,-1.75);
	\draw[fermion] (0:\radu) arc (0:90:\radu);
	\draw[fermion] (0,\radu) arc (90:180:\radu);
	\draw[vector] (180:\radu) arc (180:270:\radu); 
	\draw[vector] (270:\radu) arc (270:360:\radu); 
	\draw[dashed, dash pattern=on 5 off 3] (270:\radu) -- (270:1.5); 
	\draw[fermion] (180:\radu) -- (0:\radu);
	\draw[dashed, dash pattern=on 5 off 3] (90:\radu) -- (90:1.5); 
	\node at (-1.5, 1.75) {$\displaystyle \chi$};
	\node at ( 1.5, 1.75) {$\displaystyle \chi$};
	\node at (-1.5, -1.75) {$\displaystyle N$};
	\node at ( 1.5, -1.75) {$\displaystyle N$};
	\node at ( .5, 1.2) {$\displaystyle \varphi$};
	\node at ( .5, -1.2) {$\displaystyle h$};
	\node at ( -1.1, -.75) {$\displaystyle W$};
	\node at ( -1.1, .7) {$\displaystyle \ell_i$};
	\node at ( 1.1, .7) {$\displaystyle \ell_j$};
	\node at ( 0, .35) {$\displaystyle \nu$};
\end{tikzpicture}
	}
	&
	\raisebox{-.5\height}{
	\begin{tikzpicture}[line width=1.3 pt, scale=.8]
	\coordinate (v1) at (0,.75); 	% left vertex
	\coordinate (v2) at (0,-.75); 	% right vertex
	\coordinate (ein) at (-1,1);
	\coordinate (Zin) at (-1,-1);
	\coordinate (Zout) at (2,-1);
	\coordinate (emid) at (.5,1);
	\coordinate (eout) at (1.5,.5);
	\coordinate (phout) at (1.5,1.5);
	\coordinate (mout) at (2.5,2);
	\coordinate (eeout) at (2.5,1);
	% draw topologies
	\draw[vector] (v1)--(v2);
	\draw[fermion] (ein)--(v1);
	\draw[fermion] (v1)--(emid);
	\draw[fermion] (emid)--(eout);
	\draw[fermion] (Zin)--(v2);
	\draw[fermion] (v2)--(Zout);
	\draw[dashed, dash pattern=on 8 off 4] (emid)--(phout);
	\draw[fermion] (mout)--(phout);
	\draw[fermion] (phout)--(eeout);
%	\labels
	\node at (-1.5,1) {$\displaystyle e^-$};
	\node at (-1.5,-1) {$\displaystyle N$};
	\node at (2.5,-1) {$\displaystyle N$};
	\node at (-.5,0) {$\displaystyle \gamma$};
	\node at (1.75,.5) {$\displaystyle \mu$};
	\node at (1,1.75) {$\displaystyle \varphi$};
	\node at (2.75,2) {$\displaystyle e$};
	\node at (2.75,1) {$\displaystyle \mu$};
\end{tikzpicture}
	}
	&
	\raisebox{-.5\height}{
	\begin{tikzpicture}[line width=1.3 pt, scale = 1]
	\draw[fermionbar] (30:1)--(0,0);
	\draw[fermion] (60:1)--(0,0);
	\draw[dashed, dash pattern=on 8 off 4] (180:1)--(0,0);
	\draw[dashed, dash pattern=on 8 off 4] (-40:1)--(0,0);
	\node at (25:1.3) {$\displaystyle \ell_j$};
	\node at (55:1.3) {$\displaystyle \bar\ell_i$};
	\node at (180:1.3) {$\displaystyle h$};
	\node at (0.2,-.75) {$\displaystyle \varphi$};	
	\draw[fill=black] (0,0) circle (.07);
\begin{scope}[shift={(-40:1)}]
	\draw[fermion] (-40:1)--(0,0);
	\draw[fermionbar] (30:1)--(0,0);
	\node at (-40:1.3) {$\displaystyle \bar\ell_j$};
	\node at (30:1.3) {$\displaystyle \ell_i$};	
\end{scope}
\end{tikzpicture}
	}
	&
	\raisebox{-.5\height}{
	\begin{tikzpicture}[line width=1.3 pt, scale=.8]
	\def\labelscaling{0.4}		% distance of label from endpoint
	\def\labelscalingup{-0.45}	% distance of label from line
	\def\legangle{20}			% angle of legs
	\def\leglength{1}			% length of legs
	\def\legoffset{0.3}			% horizontal offset for long legs
	\def\lastangle{30}			% angle of final leptons
	\coordinate (v1) at (0, 1); % top vertex
	\coordinate (v2) at (0,-1); % bot vertex
	\coordinate (v1p) at (0, .75); % top vertex
	\coordinate (v2p) at (0,-.75); % bot vertex
	\coordinate (leg1a) at ($(v1)-(-\legangle:\leglength)$); 
	\coordinate (leg2a) at ($(v2)-( \legangle:\leglength)$); 
	\coordinate (leg1b) at ($(v1)+(0:\leglength)+(\legoffset,0)$);
	\coordinate (leg2b) at ($(v2)+(0:\leglength)+(\legoffset,0)$);
	\coordinate (e1a) at ($(leg1b) + (\lastangle:\leglength)$);
	\coordinate (e1b) at ($(leg1b) + (-\lastangle:\leglength)$);
	\coordinate (e2a) at ($(leg2b) + (\lastangle:\leglength)$);
	\coordinate (e2b) at ($(leg2b) + (-\lastangle:\leglength)$);
%		
%	% label positions
	\coordinate (l1a) at ($(leg1a) - \labelscaling*(-\legangle:\leglength)$); 
%	\coordinate (l1b) at ($(leg1b) + \labelscaling*(\legangle:\leglength)$);
	\coordinate (l2a) at ($(leg2a) - \labelscaling*( \legangle:\leglength)$);
%	\coordinate (l2b) at ($(leg2b) + \labelscaling*(-\legangle:\leglength)$);
	\coordinate (le1a) at ($(e1a)+\labelscaling*( \legangle:\leglength)$);
	\coordinate (le1b) at ($(e1b)+\labelscaling*(-\legangle:\leglength)$);
	\coordinate (le2a) at ($(e2a)+\labelscaling*( \legangle:\leglength)$);
	\coordinate (le2b) at ($(e2b)+\labelscaling*(-\legangle:\leglength)$);
	\coordinate (lemu) at ($(v1) !.5! (v2)$);
	\coordinate (lep1) at ($(v1) !.5! (leg1b)$);
	\coordinate (lep2) at ($(v2) !.5! (leg2b)$);
	\draw[fermion, line cap=round] (leg1a)--(v1p);
	\draw[fermion, line cap=round] (v1p)--(v2p);
	\draw[fermion, line cap=round] (v2p)--(leg2a);
	\draw[fermion, line cap=round] (e1a) -- (leg1b);
	\draw[fermion, line cap=round] (leg1b) -- (e1b);
	\draw[fermion, line cap=round] (e2a) -- (leg2b);
	\draw[fermion, line cap=round] (leg2b) -- (e2b);
	\draw[dashed, dash pattern=on 5 off 3] (v1p)--(leg1b);
	\draw[dashed, dash pattern=on 5 off 3] (v2p)--(leg2b);
%	
%	% draw nodes
	\node at (l1a) { $\displaystyle e$};
	\node at (l2a) { $\displaystyle e$};
	\node at (le1a) { $\displaystyle e$};
	\node at (le1b) { $\displaystyle \mu$};
	\node at (le2a) { $\displaystyle \mu$};
	\node at (le2b) { $\displaystyle e$};
	\node at ($(lemu) + (180:.5)$) { $\displaystyle \mu$};
	\node at ($(lep1) + (90:.5)$) { $\displaystyle \varphi$};
	\node at ($(lep2) + (270:.5)$) { $\displaystyle \varphi^*$};
\end{tikzpicture}
	}
	\\
	(a) direct detection
	& (b) $\varphi$ bremsstrahlung
	& (c) $h\to \ell_i^+\ell_i^-\ell_j^+\ell_j^-$
	& (d) $e^+e^- \to \ell_i^+\ell_i^-\ell_j^+\ell_j^-$
\end{tabular}
	\caption{Diagrams demonstrating possible search strategies at direct detection experiments, fixed target experiments, hadron colliders, and lepton colliders.
	}
	\label{fig:SM:speculate}
\end{figure}

\subsection{Direct Detection}

In the case of a flavor-conserving leptophilic mediator coupling to $\mathcal O(100~\text{ GeV})$ dark matter, the interaction with the direct detection target nuclei is loop suppressed~\cite{Kopp:2014tsa}. In our lepton flavor-violating scenarios, these interactions are further suppressed due to the flavor violating nature of the interaction. Because $\varphi$ is neutral and only couples to different-flavor leptons and the only source for lepton flavor violation in the Standard Model is the $W$ boson, the mixing between the $\varphi$ and, say, a Higgs boson only occurs at two-loop order. This is shown in Fig.~\ref{fig:SM:speculate}a. Note that this diagram is further suppressed by mass insertions of an internal charged lepton and a neutrino due to the left-handed $W$ coupling relative to the chiral $\varphi$ coupling and the GIM mechanism.

\subsection{Dark Photon Searches}

The lower mass range of $\varphi$ is comparable to the range considered by dark photon searches.
We do not include possible bounds from such searches and, instead we point unique features of light, lepton-flavor violating mediators that may pose challenges and opportunities for future experiments. 
The range of dark photon/light mediator experiments are mapped out in recent white papers on this subject, Refs.~\cite{Essig:2013lka, Alexander:2016aln}.
Of the menagerie of such experiments, the lepton-flavor violating mediator examined here is only potentially visible in a subset, 
which require some modifications.

The first requirement is that the experiment must involve leptons in order to couple to $\varphi$. Thus one is restricted to experiments with an electron beam. Because of the flavor violating couplings, $e^+e^-$ annihilation to $\varphi$ would occur through $\varphi$ pair production. This is suppressed by multiple small couplings and one cannot leverage the bump hunt strategy in $e^+e^- \to \gamma A'$. Instead, one is led to fixed target experiments that invoke $\varphi$ bremsstrahlung, $e^- N \to \mu^- N \varphi$; where $N$ is a heavy target nucleus. This is shown in Fig.~\ref{fig:SM:speculate}b. Observe that this differs from the case of a dark photon, $e^- N \to e^- N A'$, in that a muon is produced in association with the $\varphi$ as well as in the $\varphi$ decay. These searches, then, not only require higher energy electron beams, but may also have very different kinematics from the dark photon case in Ref.~\cite{Bjorken:2009mm}. One must also be careful that the searches for visible decays of a dark photon are sensitive to muons: since muons are minimum-ionizing at the typical energy scales of these experiments, one must confirm that the detectors have reasonable muon energy and/or vertex reconstruction.

For these reasons, we leave the re-interpretation of dark photon experiments to the search for lepton-flavor violating mediators for separate work. We feel that this is an interesting experimental question and may be a fruitful way to extend our search for light, weakly-coupled new physics. To the best of our knowledge, none of the experiments listed in Ref.~\cite{Alexander:2016aln} has a search for a lepton-flavor violating mediator of the type discussed here. 
The proposed ``SuperHPS'' experiment may be sensitive to $\varphi$ bremsstrahlung if the kinematics, masses, and displaced vertex resolution is amenable~\cite{Alexander:2016aln}.
We note that the SeaQuest experiment is unique in that it features the ability to accurately reconstruct muons; however, the proton initial state makes it difficult to produce the mediator~\cite{Gardner:2015wea}.

\subsection{Collider Searches}

Standard proton collider searches are limited in their reach to search for a lepton-flavor violating mediator due to the electroweak couplings required to produce progenitor leptons which may produce the mediator. That being said, these mediators can produce striking experimental signatures that are difficult to fake in the Standard Model. In this case, it is useful to invoke the electroweak-scale, non-renormalizable Lagrangian interaction (\ref{eq:L:phi:SM:EW}) that was assumed to generate the low-energy couplings (\ref{eq:L:phi:SM}). From this interaction, one can consider Higgs decays such as $h\to \varphi^* \mu \bar e \to \mu\bar\mu e\bar e$. This is shown in Fig.~\ref{fig:SM:speculate}c. Note that with the chiral structure imposed, one cannot produce Higgs decays to two pairs of same-sign, same-flavor leptons unless $\varphi$ is a real field.
The possibility of an exotic flavor violating Higgs decay mediated by new scalars has recently generated attention~\cite{Dery:2014kxa, Baek:2015fma, Liu:2015oaa} in part due to the possibility of an observation of $h\to \tau \mu$ at the Large Hadron Collider~\cite{Khachatryan:2015kon, Aad:2015gha}, though those claims appear to be in tension with early results from Run II~\cite{CMS:2016qvi,Aad:2016blu}. We postpone a discussion to future work~\cite{Iftahworkinprogress}.

While the ideal collider search would require a muon-electron collider~\cite{Choi:1997bm}, low-energy $e^+e^-$ colliders like the $B$-factories may be sensitive to lepton-flavor violating mediators. The upcoming Belle-II run, for example, will run at $\sqrt{s} = 10~\text{GeV}$ and is sensitive to muon final states~\cite{TheBelle:2015mwa}. 
The symmetry structure dictates that if $\varphi$s are produced,
the leading signals involve an even number of opposite-sign
same-flavor lepton pairs.
For example, one may search for 
$e^+e^- \to  e\bar e \mu\bar \mu$.
This process has contributions from $\varphi\varphi^*$ pair production, in which each opposite-flavor-opposite-charge pair reconstructs a $\varphi$, as shown in Fig.~\ref{fig:SM:speculate}d,
or at higher order through a flavor changing $\varphi$-strahlung off 
an initial $e^+$ state.
Similar symmetry arguments apply for dark matter searches in possible future high-energy $e^+e^-$ colliders. In this case, the production of dark matter is $\varphi$-mediated,
so the leading contributions are:
(a) $e^+e^-\to \chi\bar\chi\chi\bar\chi$, 
which is suppressed at order $g_{e\mu}^4$ and its reach
is limited by phase-space due to the multiple
dark matter particles,
and
(b) $e^+e^-\to \chi\bar\chi\varphi, \varphi\to e\mu$.
which is also $g_{e\mu}^4$ suppressed.

\section{Conclusions \& Outlook}

\begin{table} 
	\renewcommand{\arraystretch}{1.3} % spacing between rows
	\setlength{\tabcolsep}{12pt}
	\centering
	\begin{tabular}{ @{} cccccc @{} } \toprule
		Dark parity of $\varphi$
		& Annihlation 
		& $\ell_i\ell_j$
		& $m_\chi$/GeV
		& %$\langle\sigma v\rangle/\langle\sigma v\rangle_\text{rel.}$
		$\displaystyle\frac{\langle\sigma v\rangle\phantom{_\text{rel.}}}{\langle\sigma v\rangle_\text{rel.}}$
		& AMS-02	
	\\ \hline
		\multirow{3}{*}{%
		Not parity eigenstate
		}
		& \multirow{3}{*}{\raisebox{-.5\height}{\begin{tikzpicture}[line width=1 pt, scale=.6]
	\def\labelscaling{0.4}		
	\def\labelscalingup{-0.45}	
	\def\legangle{20}			% angle of legs
	\def\leglength{.9}			% length of legs
	\def\legoffset{0.3}			% horizontal offset for long legs
	\def\labsize{}				% e.g. \large
	\coordinate (v1) at (0, .75); % top vertex
	\coordinate (v2) at (0,-.75); % bot vertex
	\coordinate (leg1a) at ($(v1)-(-\legangle:\leglength)$); 
	\coordinate (leg2a) at ($(v2)-( \legangle:\leglength)$); 
	\coordinate (leg1b) at ($(v1)+( \legangle:\leglength)+(\legoffset,0)$);
	\coordinate (leg2b) at ($(v2)+(-\legangle:\leglength)+(\legoffset,0)$);

	\draw[fermion, line cap=round] (leg1a)--(v1);
	\draw[fermion, line cap=round] (v1)--(v2);
	\draw[fermion, line cap=round] (v2)--(leg2a);
	\draw[dashed, dash pattern=on 5 off 3] (v1)--(leg1b);
	\draw[dashed, dash pattern=on 5 off 3] (v2)--(leg2b);
\end{tikzpicture}}}
		& $\tau e$
		& $20-40$
		& $1\, (0.4)$ 
		& \textcolor{green!50!black}{\checkmark}
	\\
		& 
		& $\tau \mu$
		& $20-40$
		& $1\, (0.4)$ 
		& \textcolor{green!50!black}{\checkmark}
	\\
		&
		& $\mu e$ 
		& $30-90$
		& $3\, (1.1)$  
		& \textcolor{red!75!black}{\ding{55}} (\textcolor{green!50!black}{\checkmark})
	\\ 
%	\hline
\hdashline
		\multirow{3}{*}{Pseudoscalar (parity-odd)}
		& \multirow{3}{*}{\raisebox{-.5\height}{\begin{tikzpicture}[line width=1 pt, scale=.6]
	\def\labelscaling{0.4}		% distance of label from endpoint
	\def\labelscalingup{-0.45}	% distance of label from line
	\def\legangle{20}			% angle of legs
	\def\leglength{.9}			% length of legs
	\def\legoffset{0.3}			% horizontal offset for long legs
	\def\labsize{}
	\coordinate (v1) at (0, 1); % top vertex
	\coordinate (v2) at (0,-1); % bot vertex
	\coordinate (v3) at (0, 0); % middle vertex
	\coordinate (leg1a) at ($(v1)-(-\legangle:\leglength)$); 
	\coordinate (leg2a) at ($(v2)-( \legangle:\leglength)$); 
	\coordinate (leg1b) at ($(v1)+( \legangle:\leglength)+(\legoffset,0)$);
	\coordinate (leg2b) at ($(v2)+(-\legangle:\leglength)+(\legoffset,0)$);
	\coordinate (leg3b) at ($(v3)+(0:\leglength)+(\legoffset,0)$);
	\draw[fermion, line cap=round] (leg1a)--(v1);
	\draw[fermion, line cap=round] (v1)--(v3);
	\draw[fermion, line cap=round] (v3)--(v2);
	\draw[fermion, line cap=round] (v2)--(leg2a);
	\draw[dashed, dash pattern=on 5 off 3] (v1)--(leg1b);
	\draw[dashed, dash pattern=on 5 off 3] (v2)--(leg2b);
	\draw[dashed, dash pattern=on 5 off 3] (v3)--(leg3b);
\end{tikzpicture}}}
		& $\tau e$
		& $20-40$
		& $1\, (0.4)$ 
		& \textcolor{green!50!black}{\checkmark}
	\\
		& 
		& $\tau \mu$
		& $20-40$
		& $1\, (0.4)$ 
		& \textcolor{green!50!black}{\checkmark}	\\
		&
		& $\mu e$ 
		& $40-100$
		& $4\, (1.4)$ 
		& \textcolor{red!75!black}{\ding{55}} (\textcolor{green!50!black}{\checkmark})
		\\ \bottomrule
	\end{tabular}
	\caption{
		Summary of Figs.~\ref{fig:GCEspectra_parameterized} and \ref{fig:GCEspectra_sys_eng_bin}. The annihilation cross-section, $\langle \sigma v\rangle$ is given for $\gamma_\text{\tiny NFW} = 1.0$ (1.2). The $\mu e$ modes are in tension with the AMS-02 positron bound unless one takes the contracted $\gamma_\text{\tiny NFW} = 1.2$ profile.
		The range of mediator masses are given in~(\ref{eq:mass:ranges}).
		\label{table:summary}
	}
	
\end{table}

In this manuscript we examined a class of models where dark matter interacts with the Standard Model through spin-0 mediators with chiral, flavor violating interactions to leptons. For a range of mediator masses, this set up realizes the secluded dark matter scenario where the relic abundance and indirect detection annihilation rates are controlled by one set of couplings, while direct detection, collider bounds, and low-energy searches are controlled by a separate set of couplings. 

We have shown that in the dark sector, one is able to simultaneously achieve a thermal relic and the observed Fermi $\gamma$-ray excess without causing tension with the measured AMS-02 positron spectrum.
The $\gamma$-ray excess is produced through a combination of prompt photon emission for $\tau$ final states and the inverse Compton scattering of the interstellar radiation field for $\mu$ and $e$ final states.
Because dwarf spheroidals have much weaker interstellar radiation fields, this helps alleviate tensions of the $\gamma$-ray excess with non-observations in dwarf spheroidals.
For the case of $\mu e$ interactions, however, this requires a contracted dark matter halo profile. The spectra of the Standard Model byproducts of dark matter annihilation are softened because the decay goes through on-shell mediators.  This smearing helps the $e^+e^-$ spectrum to fit within the error bars of the AMS experiment. We have commented that the parameters for the $\mu e$ final state appear to be consistent with the target region for a self-interacting dark matter solution to small scale structure anomalies. The dark sector interactions are summarized in Table~\ref{table:summary}.

One unique feature of the chiral lepton-flavor violating interactions is that the bounds on the Standard Model couplings are weaker than direct flavor-conserving interactions. 
We have shown that the upper bounds on the Standard Model interactions in this scenario come from the forward--backward asymmetry in $e^+e^- \to f\bar f$ and from the anomalous magnetic moments of the muon and electron. For the case of a $\tau\mu$ interaction, one can simultaneously explain the $(g-2)_\mu$ anomaly. We have shown that the chiral flavor structure of the mediator--Standard Model interaction suppresses bounds from conventional charged lepton flavor violation experiments and direct detection experiments. 
We explained that these suppressions are straightforward to understand from the point of view of a spurious $L_i - L_j$ symmetry that is respected by the mediator when it is complex. Finally, we point out possible opportunities in dark photon experiments and collider searches that are motivated by this class of mediator models. 

We emphasize that while we have benchmarked our models for the Fermi $\gamma$-ray excess, the class of models are independently meaningful as an example of light, weakly-coupled new physics that can play an important role in both Standard Model and dark matter phenomenology and that are able to avoid current constraints.

\section*{Acknowledgements}

We enthusiastically thank 
Sheldon Campbell, 
Jared Evans,
Paddy Fox,
Manoj Kaplinghat,
G\o rd\aa n Kr$n$j\aa i\c{c},
Arvind Rajaraman
Yael Shadmi,
Tim M.~P.~Tait,
Scott Thomas,
Hai-Bo Yu,
and 
Yue Zhao
for many useful comments and discussions. We especially thank Jared for discussions regarding the contribution to anomalous magnetic moments and possible collider searches at high luminosity, Paddy for pointing out the possible relation to $h\to \tau\mu$, and Yael for suggesting that a complex mediator can be treated as a spurion for $L_i-L_j$ breaking.
\textsc{a.k.} is supported by an \textsc{nsf} graduate research fellowship.
\textsc{p.t.} thanks the Aspen Center for Physics (\textsc{nsf} grant \#1066293) for its hospitality during a period where part of this work was completed and sends hugs to Jane Kelly and Patty Fox.
This work is supported in part by the \textsc{nsf} grant \textsc{phy}-1316792 and \textsc{nsf} grant-1620638.

\bibliographystyle{utphys} 
\bibliography{LFVmedbib}

\end{document}